\documentclass{aa}  

\usepackage{graphicx}
\usepackage{bm}
\usepackage{xcolor}
\usepackage{nccmath}

\usepackage[capitalize]{cleveref}
\crefname{section}{Sect.}{Sects.}
\Crefname{section}{Section}{Sections}
\usepackage{multirow}

\newcommand{\ev}[1]{\left\langle #1 \right\rangle}

\newcommand{\e}[1]{_{\text{#1}}}

\newcommand{\dd}{\mathrm{d}}

\usepackage{txfonts}

\begin{document} 

   \title{Impact of lensing magnification on the analysis of galaxy clustering in redshift space}
   \titlerunning{Impact of lensing magnification on RSD}

   \author{Michel-Andrès Breton \inst{1,2,3}\thanks{\email{breton@ice.csic.es}}
            \and Sylvain de la Torre \inst{1}
            \and Jade Piat \inst{1}
            }
            
   \institute{Aix Marseille Univ, CNRS, CNES, LAM, Marseille, France
                \and
                Institute of Space Sciences (ICE, CSIC), Campus UAB, Carrer de Can Magrans, s/n, 08193 Barcelona, Spain
                \and
                Institut d’Estudis Espacials de Catalunya (IEEC), Carrer Gran Capit\`a 2-4, 08193 Barcelona, Spain               
                }
   
   \date{Received ---; accepted ---}
 
    \abstract{We study the impact of lensing magnification on the observed three-dimensional galaxy clustering in redshift space. We used the RayGal suite of $N$-body simulations, from which we extracted samples of dark matter particles and haloes in the redshift regime of interest for future large redshift surveys. Several magnitude-limited samples were built that reproduce various levels of magnification bias ranging from $s=0$ to $s=1.2$, where $s$ is the logarithmic slope of the cumulative magnitude number counts, in three redshift intervals within $1<z<1.95$. We studied the two-point correlation function multipole moments in the different cases in the same way as would be applied to real data, and investigated how well the growth rate of structure parameter could be recovered. In the analysis, we used an hybrid model that combines non-linear redshift-space distortions and linear curved-sky lensing magnification. We find that the growth rate is underestimated when magnification bias is not accounted for in the modelling. This bias becomes non-negligible for $z>1.3$ and can reach 10\% at $z\simeq 1.8$, depending on the properties of the target sample. In our data, adding the lensing linear correction allowed us to recover an unbiased estimate of the growth rate in most cases when the correction was small, even when the fiducial cosmology was different from that of the data.  For larger corrections (high redshifts, low bias, and high $s$ value), we find that the weak-lensing limit has to be treated with caution as it may no longer be a good approximation . Our results also show the importance of knowing $s$ in advance instead of letting this parameter free with flat priors because in this case, the error bars increase significantly.}
 

   \maketitle
%

\section{Introduction}


The observation of distant galaxies provides a wealth of information regarding the nature and content of our Universe. Since galaxies trace the underlying matter density field, their spatial distribution can be used to probe the evolution of the large-scale structure. Moreover, the Doppler effect induced by the peculiar motion of galaxies, also known as redshift-space distortions \citep[RSD;][]{kaiser1987clustering}, leaves a distinct imprint on the three-dimensional clustering of these distant galaxies, which can in turn be used to gain information on the linear growth rate of structure. The latter is directly sensitive to the theory of gravity \citep{guzzo2008}. 

Galaxy clustering has been used for more than two decades to place constraints on the $f\sigma_8$ parameter, that is, the growth rate multiplied by the power spectrum normalisation. No significant deviations from General Relativity were found \citep[e.g.][]{peacock2001measurement, tegmark2006cosmological,blake2012wigglez,delatorre2013vipers,alam2017clustering, alam2021completed}.
Future surveys such as Euclid \citep{laureijs2011euclid}, the Subaru Prime Focus Spectrograph \citep{takada2014extragalactic}, the Square Kilometre Array \citep{ska2020cosmology}, or the recently started DESI \citep{desi2016desi} will probe the distribution of galaxies with unprecedented accuracy and at higher redshifts ($z > 1$) than before.
In particular, Euclid will provide a $1-3\%$ level of precision on $f\sigma_8$ between $z = 0.9$ and $z=1.8$ \citep{laureijs2011euclid,majerotto2012probing}.

Because future galaxy surveys will probe higher redshifts with high accuracy, the theoretical modelling of galaxy clustering in the form of summary statistics such as the two-point correlation function will need to be improved to provide an unbiased estimate of cosmological parameters. More precisely, the current modelling of the observed two-point correlation function (or power spectrum) only accounts for peculiar velocities (RSD), while analytical works have shown in linear theory that the next dominant term at high redshift will be due to gravitational lensing \citep{yoo2009new,challinor2011linear,bonvin2011what}, also known as magnification bias \citep{schneider1992gravitational}. Gravitational lensing indeed modifies the apparent angular position of sources due to the deflection of light rays along their path, and it also magnifies their fluxes, so that in magnitude-limited surveys, we can find more or fewer objects in magnified regions, depending on the slope of the luminosity function.

It is interesting to remark that in galaxy-galaxy lensing, the theoretical framework required to incorporate the magnification bias corrections was first devised in \cite{ziour2008magnification}, but this gained more attention only very recently \citep{duncan2014complementarity,ghosh2018general,thiele2020disentangling,unruh2020importance,joachimi2021kids1000,lee2022probing}. The difference in theoretical modelling between Dark Energy Survey one-year \citep{prat2018des1yrGGL} and three-year \citep{des2022des3yrGCWL} galaxy-galaxy lensing analyses, where only the latter accounts for the effect of lensing magnification, is particularly telling. Similarly, several analytical works have investigated the effect of magnification bias on the three-dimensional two-point correlation function \citep{matsubara2000gravitational,hui2007anisotropic,hui2008anisotropic}, but has not been implemented in observational analysis so far. The main reason for this is that, until now, galaxy redshift surveys have probed low enough redshifts, where gravitational lensing does not have an important impact on the overall three-dimensional distribution of galaxies. However, in the near future, this may no longer be the case. Of particular relevance for the present paper, \cite{jelic-cizmek2021importance} investigated the effect of lensing magnification in spectroscopic surveys and found that neglecting magnification bias leads to an underestimation of the growth rate of structure. This work was restricted to pure analytical models in linear theory and in a Fisher analysis. From a numerical point of view, \cite{elkhashab2022large} assessed the detectability of magnification bias on the power spectrum monopole only for a Euclid-like survey.

The present paper aims at investigating the impact of magnification bias on the determination of the growth rate from clustering in redshift space in a typical observational set up. We use a suite of $N$-body simulations that accounts for the fully non-linear structure formation and perform realistic galaxy clustering analyses, similarly as in observations, in different regimes of magnification bias. We investigate a minimal model to account for the magnification effect on the multipole moments of the redshift-space correlation function, and we study the accuracy with which the growth rate of structure parameter can be recovered.

The paper is organised as follows. In \Cref{sec:theory} we discuss the theoretical modelling of the magnified correlation function multipole moments in redshift space. The numerical data and method we used to perform a likelihood analysis are presented in \Cref{sec:methods}. Finally, the results are shown in \Cref{sec:results}, and we conclude in \Cref{sec:conclusion}.

\section{Theory}
\label{sec:theory}

The observed spatial distribution of galaxies in the Universe depends on two main aspects: first, galaxies are biased tracers of the matter density field, which means that in `real space' (i.e. the Universe as it is really), galaxies are located at local matter density peaks
formed through gravitational instabilities. Second, the observation of galaxies through their emitting light modifies our perception of their distribution. In particular, their measured redshift can be shifted with respect to their Friedmann-Lemaitre-Robertson-Walker counterpart due to peculiar velocities (RSD) and due to local or integrated gravitational potentials. This leads to distortions when estimating the distance to galaxies given a fiducial cosmology, and therefore modifies their overall perceived spatial distribution. Moreover, light rays follow null geodesics and can therefore be deflected by gravitational potentials along their trajectories, which impacts the observed angular position of sources. In the following, we present how the two dominant effects leading to the observed redshift-space galaxy two-point correlation function, that is, RSD and gravitational lensing magnification, can be modelled.

\subsection{Redshift-space distortions}
\label{subsec:rsd_modelling}

The standard way of modelling clustering in redshift space beyond linear theory \citep{peebles1980LSS,kaiser1987clustering,hamilton1992measuring} is to consider only redshift perturbations due to peculiar velocities. The real-space quasi-linear clustering can be predicted with different flavours of perturbation theory, either Eulerian \citep{bernardeau2002large,crocce2006renormalized,taruya2012direct, taruya2013precision} or Lagrangian \citep{zeldovich1970gravitational,matsubara2008nonlinear,matsubara2008resumming,carlson2013convolution}. In addition, the mapping from real to redshift space can be done with different approaches. Common approaches in observational studies include the TNS model \citep{taruya2010baryon,taruya2013precision} and the streaming model \citep{scoccimarro2004redshift,reid2011towards}.

In this work, we consider the \emph{\textup{convolution Lagrangian perturbation theory}} \citep[CLPT;][]{carlson2013convolution} to predict real-space quantities. Generally, if the non-linear clustering in real space is well modelled with this approach, Lagrangian theories are not accurate enough on small scales in redshift space \citep{white2014zeldovich}. We therefore adopt the Gaussian streaming model \citep{reid2011towards} for the
mapping from real to redshift space. The joint use of CLPT and Gaussian streaming (CLPT-GS) has been shown to provide a good match to data and simulations \citep{wang2014analytic} and is routinely used to model observations
\citep[e.g. ][]{reid2012the,samushia2014the,zarrouk2018clustering,bautista2021completed}.

\subsubsection{Convolution Lagrangian perturbation theory}
\label{subsubsec:CLPT}

In the Lagrangian framework, the position $\bm{x}$ of an infinitesimal volume element is given by
\begin{equation}
 \bm{x}(\bm{q},t) = \bm{q} + \bm{\Psi}(\bm{q}, t),   
\end{equation}
where $\bm{q}$ is the Lagrangian coordinate (initial position) and $\bm{\Psi}(\bm{q})$ the displacement field. The latter encodes the displacement of any mass element. For conciseness, in the following we omit its time dependence. In Lagrangian theories, the displacement field is assumed to be small, whereas for Eulerian theories, the density contrast is assumed to be small. The displacement field can thus be expanded as a perturbative series, $\bm{\Psi} = \sum_{n=1}^\infty \bm{\Psi}^{(n)}$, where the first-order solution is the well-known Zel'dovich approximation \citep{zeldovich1970gravitational}.

Due to mass conservation, the relation between the density field of a volume element depends on its initial location through
\begin{equation}
\left[1+\delta(\bm{x})\right]\dd^3\bm{x} = \left[1+\delta(\bm{q})\right]\dd^3\bm{q},
\end{equation}
where $\delta$ is the matter density contrast. We have therefore that
\begin{equation}
    1+\delta(\bm{x}) = \int \dd^3\bm{q} ~ \delta_D(\bm{x}-\bm{q}-\bm{\Psi}),
    \label{eq:matter_density}
\end{equation}
where $\delta_D$ is the Dirac delta function. However, the  matter density field is not observationally relevant for galaxy clustering, but we are rather interested in biased tracers of the underlying matter field. By assuming a local Lagrangian bias, we can write
\begin{equation}
    1 + \delta_g(\bm{q}) = F(\delta(\bm{q})),
\end{equation}
where $\delta_g$ is the density contrast of galaxies and $F(\delta)$ is the biasing function smoothed on a given scale. We note that at first order, the linear Eulerian and Lagrangian biases are simply related by $b_E^{(1)}= 1 + b_L^{(1)}$. In the following, we use the notation $b_n = b_L^{(n)}$, with $b_1 = \ev{F'}$ and $b_2=\ev{F''}$ the first- and second-order Lagrangian bias parameters given by the expectation value of the first- and second-order derivatives of $F(\delta)$ respectively.

The counterpart of \cref{eq:matter_density} for biased tracers is thus
\begin{equation}
  1 + \delta_g(\bm{x}) = \int \dd^3\bm{q} ~ F[\delta(\bm{q})] ~ \delta_D(\bm{x}-\bm{q}-\bm{\Psi}).
\end{equation}
From this expression, it is possible to derive the real-space galaxy two-point correlation function, mean pairwise velocity, and velocity dispersion as
\begin{eqnarray}
    \label{eq:xi_realspace}
1 + \xi(\bm{r}) &=& \ev{\left[1+\delta_g(\bm{x})\right]\left[1+\delta_g(\bm{x}+\bm{r})\right]} \nonumber \\
               &=& \int \dd^3\bm{q} ~ M_0(\bm{r}, \bm{q}), \\
    \label{eq:v12_realspace}
    v_{12}(\bm{r}) &=& \frac{\ev{\left[1+\delta_g(\bm{x})\right]\left[1+\delta_g(\bm{x}+\bm{r})\right]\left[ v_z(\bm{x}+\bm{r})-v_z(\bm{x})\right]}}{\ev{\left[1+\delta_g(\bm{x})\right]\left[1+\delta_g(\bm{x}+\bm{r})\right]}} \nonumber \\
              &=& [1+\xi(r)]^{-1}\int \dd^3\bm{q} ~ M_1(\bm{r}, \bm{q}), \\
    \label{eq:s12_realspace}
    \sigma_{12}^2(\bm{r}) &=& \frac{\ev{\left[1+\delta_g(\bm{x})\right]\left[1+\delta_g(\bm{x}+\bm{r})\right]\left[ v_z(\bm{x}+\bm{r})-v_z(\bm{x})\right]^2}}{\ev{\left[1+\delta_g(\bm{x})\right]\left[1+\delta_g(\bm{x}+\bm{r})\right]}} \nonumber \\
    &=& [1+\xi(r)]^{-1}\int \dd^3\bm{q} ~ M_2(\bm{r}, \bm{q}),
\end{eqnarray}
where $r = |\bm{r}|$ is the scale, $v_z$ is the velocity along the line of sight $\bm{u}_z$ with $v_z(\bm{x}_2)-v_z(\bm{x}_1) = (\dot{\bm{x}}_2-\dot{\bm{x}}_1)\cdot\bm{u}_z$, and $M_0$, $M_1$ and $M_2$ are integration kernels containing the parameters of the model \citep{wang2014analytic}. The main difference between CLPT and other Lagrangian perturbation theories such as in \citet{matsubara2008resumming} lies in the resummation employed in the kernels.

\subsubsection{Gaussian streaming model}
\label{subsubsec:GS}

The CLPT predictions for the real-space correlation function and the two first moments of the pairwise velocity distribution, that is, \cref{eq:xi_realspace,eq:v12_realspace,eq:s12_realspace}, can be used in the Gaussian streaming model \citep{reid2011towards} to predict the anisotropic two-point correlation function in redshift space. In this model, the redshift-space correlation function is recovered by convolving the real-space correlation function along the line of sight, with a scale-dependent pairwise velocity distribution taken to be Gaussian. This gives
\begin{multline}
    1 + \xi^s(r_\perp, r_\parallel) = \int \frac{\dd y}{\sqrt{2\pi\sigma^2(r,\upsilon)}} [1+\xi(r)] \\
       \times \exp\left\{-\frac{\left[r_\parallel - y - \upsilon v_{12}(r)\right]^2}{2\sigma^2(r,\upsilon)}\right\},
       \label{eq:gaussian_streaming}
\end{multline}
where $r_\parallel$ and $r_\perp$ are the components of the separation vector parallel and perpendicular to the line of sight, $r^2 = r_\perp^2 + y^2$,  $\upsilon = y/r$ is the end-point line-of-sight cosine angle definition, and $\sigma^2(r,\upsilon) = \sigma_{12}^2(r,\upsilon)+\sigma^2_v$, with $\sigma_v$ an additional velocity dispersion term that accounts for small-scale random motions in virialised objects. From \cref{eq:gaussian_streaming}, it is fairly easy to compute the multipole moments of the correlation function by integrating over the Legendre polynomials. We note that one limitation of the Gaussian streaming model as presented here is that it assumes the distant-observer approximation, meaning that it neglects wide-angle effects \citep{szalay1998redshift,szapudi2004wide,papai2008nonperturbative,raccanelli2010simulating,reimberg2016redshift,castorina2018zeldovich,beutler2019interpreting,taruya2020wide}. In our case, we restrict our analysis to relatively high redshifts ($z > 1$) so that we can safely work under this approximation. Our implementation of the CLPT-GS model is publicly available\footnote{\url{https://github.com/mianbreton/CLPT_GS}}.

\subsection{Magnification bias}
\label{subsec:magnification_bias}

Beyond the redshift-space distortions induced by peculiar velocities, the next dominant effect at $z>1$ that modifies the apparent distribution of galaxies is magnification bias. It originates from the fact that some regions of the sky are magnified (demagnified) due to gravitational lensing, meaning that they contain fewer (more) sources than on average. If we consider a survey of galaxies selected in flux, gravitational lensing will modify the apparent fluxes of the galaxies, so that some of them will enter or exit the sample. This selection effect will impact the observed clustering of galaxies. It is worth noting that magnification bias is not the only effect that arises: additional effects ensue that depend on peculiar velocities or gravitational potential \citep{challinor2011linear}. However, at high redshift, magnification bias is the dominant effect and is the focus of this work.

\subsubsection{Impact on the number counts}
\label{subsubsec:lensed_number_counts}
The surface brightness of sources, defined as the flux per unit solid angle, is conserved through gravitational lensing. This implies that the apparent size and flux of a source are simultaneously modified as
\begin{eqnarray}
    S'(\bm{\theta}) &=& \mu(\bm{\theta}) S(\bm{\theta}) , \\
    \dd\Omega'(\bm{\theta}) &=& \mu(\bm{\theta}) \dd\Omega(\bm{\theta}),
\end{eqnarray}
where $S$, $\mu,$ and $\dd\Omega$ are the flux, magnification, and solid angle, respectively, in the direction $\bm{\theta}$ on the sky (in the following we omit the angular dependence), and a prime denotes a magnified quantity. The conservation of the number of sources can be written as
\begin{equation}
n'(m')\dd m'\dd\Omega' = n(m)\dd m\dd\Omega,
\end{equation}
where $n(m)$ is the number density of sources per unit of solid angle and per magnitude interval $dm$,  $m = -2.5\log(S) + C$ is the magnitude, and $C$ is a constant. From this, it is straightforward to infer the magnified magnitude $m' = m - 2.5\log(\mu)$. The total number of observed sources up to a given magnitude limit $m_l$ is 
\begin{eqnarray}
 \int_{-\infty}^{m_l} n'(m')\dd m' &=& \mu^{-1}\int_{-\infty}^{m_l} n\left(m'+2.5\log(\mu)\right)\dd m' \\
                         &=& \mu^{-1} \int_{-\infty}^{m_l+2.5\log(\mu)} n(m) \dd m .
\end{eqnarray}
Ultimately, this can be rewritten in terms of cumulative number densities as
\begin{equation}
    n'(<m_l) = \mu^{-1} n\left(<m_l+2.5\log(\mu)\right) .
    \label{eq:cumulative_general}
\end{equation}
One generally considers a simple model for the luminosity function, a power law such that \citep{schneider1992gravitational,broadhurst1995mapping}\begin{equation}
    n(<m) \propto 10^{ms} ,
    \label{eq:power_law}
\end{equation}
where $s$ is defined as the logarithmic slope of the cumulative distribution and is a property of the target sample. We note that for this specific function, cumulative and differential distributions have the same shape. By inserting \cref{eq:power_law} in \cref{eq:cumulative_general}, we obtain that
\begin{eqnarray}
  \label{eq:delta_lens}
  \Delta\e{len} &=& \mu^{2.5s-1} - 1 \\
  \label{eq:delta_lens_approx}
             &\approx& (5s-2) \kappa,
\end{eqnarray}
where $\Delta\e{len} = n'(<m_l)/n(<m_l) - 1$ is the perturbation on the number count due to magnification bias in a given direction on the sky, and $\kappa$ is the lensing convergence. In the second line we have performed a first-order Taylor expansion on the magnification as $\mu = 1 + 2\kappa$ (hereafter referred to as the `weak-lensing limit'), with $|\kappa|\ll 1$. Gravitational lensing induces two competing terms on the observed number counts as seen in \cref{eq:delta_lens_approx}: the $s$-dependent term implies that in magnified regions ($\mu>1$, $\kappa>0$), the flux of sources increases so that we are more likely to find objects (and conversely, the chances of finding objects in demagnified regions are lower). The second term, which is usually referred to as `dilution bias', describes the change in size of solid angles on the sky. Magnified regions occupy more space, so that for a constant density, we should find fewer objects in these regions than on average. These two effects cancel exactly for $s = 0.4$.

\subsubsection{Two-point correlation function correction}

To be consistent, the lensing correction associated with magnification bias on the correlation function should in principle be derived using the same theoretical framework as for RSD. However, these developments are beyond the scope of the present paper. We instead propose a simple correction based on linear theory, which can easily be used in addition to any RSD model. 

We start from the observed galaxy number counts, which accounts for density, RSD, and lensing perturbations (the full expressions accounting for all the terms at first order in metric perturbations can be found in \citealt{yoo2009new}, \citealt{challinor2011linear} and \citealt{bonvin2011what}),
\begin{equation}
\label{eq:total_number_counts}
    \Delta = \Delta\e{den} + \Delta\e{rsd} + \Delta\e{len},
\end{equation}
where $\Delta\e{den} = b\delta$, $b$ is the Eulerian linear bias, and $\Delta\e{rsd}=-\partial_r v_r/\mathcal{H}$ is the RSD component, where $\partial_r v_r$ and $\mathcal{H}$ are the gradient of the velocity field along the line of sight and the conformal Hubble parameter, respectively. The lensing perturbation $\Delta\e{len}$ is that of \cref{eq:delta_lens_approx}. We note that the decomposition in \cref{eq:total_number_counts} is only true at first order since it neglects higher-order lensing correlations.

Since the correlation function can be written as $\xi(\bm{r}) = \ev{\Delta(\bm{x})\Delta(\bm{x}+\bm{r})}$, the linear correction that comes from the addition of lensing magnification in the number counts is
\begin{equation}
\label{eq:lensing_correction_xi}
\xi\e{corr}(\bm{r}) = \xi\e{den-len}(\bm{r}) + \xi\e{rsd-len}(\bm{r}) + \xi\e{len-len}(\bm{r}),
\end{equation}
where $\xi\e{A-B}(\bm{r}) \equiv \ev{\Delta\e{A}(\bm{x})\Delta\e{B}(\bm{x}+\bm{r})}$. The expressions for the different terms in \cref{eq:lensing_correction_xi} are derived in  \cite{matsubara2000gravitational}, \cite{hui2007anisotropic}, \cite{hui2008anisotropic}, and in \cite{tansella2018fullsky}, \cite{tansella2018coffe} for the curved-sky case. Precisely, in the latter case we have
\begin{equation} \label{eq:xiab}
    \xi\e{A-B}(\theta, z_1, z_2) = \int \frac{\dd k}{k} P_{\rm R}(k) \, \mathcal{Q}_k^{\rm A-B}(\theta, z_1, z_2), 
\end{equation}
where $(\theta, z_1, z_2)$ defines the separation vector in observed coordinates\footnote{Here, $z_1$ and $z_2$ are the redshifts of the objects of the pair, and $\theta$ is the angle between them.}, $P_{\rm R}(k)$ is the primordial matter power spectrum, and the kernels $\mathcal{Q}_k^{\rm A-B}$ with $ \textrm{A-B} = \{\textrm{den-len}, \textrm{rsd-len}, \textrm{len-len}\}$ read
\begin{align}
    \mathcal{Q}_k^\text{den-len}(\theta,z_1,z_2) &= b(z_1) S_D(z_1) \left(\frac{2 - 5 s}{2\chi_2}\right) \\ \notag & \int_0^{\chi_2} \dd \lambda  \frac{\chi_2-\lambda}{\lambda} S_{\phi+\psi}(\lambda)  \, \zeta^{0L}(k\chi_1,k\lambda,\theta)   \,,\\
    \mathcal{Q}_k^\text{rsd-len}(\theta,z_1,z_2) &=  \frac{k}{\mathcal{H}(z_1)}S_V(z_1) \left(\frac{2 - 5 s}{2\chi_2}\right) \\ \notag & \int_0^{\chi_2} \dd \lambda  \frac{\chi_2-\lambda}{\lambda}S_{\phi+\psi}(\lambda) \, \zeta^{2L}(k\chi_1,k\lambda,\theta)  \,, \\
    \mathcal{Q}_k^\text{len-len}(\theta,z_1,z_2) &=\!  \frac{\left(2 - 5 s\right)^2}{4 \chi_1\chi_2} \! \\ \notag & \int_0^{\chi_1} \!\! \!\!  \int_0^{\chi_2} \!\!\!\dd \lambda \dd \lambda'  \frac{(\chi_1\!-\!\lambda)(\chi_2\!-\!\lambda')}{\lambda\lambda'}S_{\phi+\psi}(\lambda)\\
    & S_{\phi+\psi}(\lambda') \zeta^{LL}(k\lambda,k\lambda',\theta). \notag
\end{align}
In these equations, $\zeta$ are pure geometrical functions provided in Appendix B of \citet{tansella2018fullsky}, $\chi_i$ is the comoving distance to redshift $z_i$, and $S_{D}$, $S_{V}$, and $S_{\phi+\psi}$ are the scaled transfer functions associated with density, peculiar velocity, and gravitational potentials, respectively. $\xi\e{A-B}(\theta, z_1, z_2)$ in \cref{eq:xiab} can be written in terms of the separations parallel and perpendicular to the line of sight, $r_\parallel$ and $r_\perp$, using that $r_\parallel=\chi_2-\chi_1$, $r_\perp=\sqrt{r^2 - r^2_\parallel}$, and $r=\sqrt{\chi^2_1 +\chi^2_2 - 2\chi_1 \chi_2 \cos \theta}$.

We implement the magnification bias correction using the \textsc{Coffe} library\footnote{\url{ https://github.com/JCGoran/coffe}} \citep{tansella2018coffe}, which directly provides $\xi\e{den-len}$, $\xi\e{rsd-len}$, $\xi\e{len-len}$ in bins of $(r_\perp, r_\parallel)$ using curved-sky linear theory and given an input linear power spectrum. \cite{jelic-cizmek2021flatsky} noted that although there is no large differences between the curved-sky and flat-sky prescriptions at the scales of interest for us in general, that is $r \lesssim 150~h^{-1}$Mpc, the $\xi\e{den-len}$ component is quite sensitive to the adopted prescription. We therefore adopted the full curved-sky implementation. 

Our theoretical model for the redshift-space correlation function therefore consists of the CLPT-GS prediction for non-linear RSD and of the curved-sky linear theory prediction for the additional lensing magnification correction. Formally, the anisotropic correlation function model is given by
\begin{equation}
    \xi\e{model}(r_\perp, r_\parallel) = \xi\e{CLPT-GS}(r_\perp, r_\parallel)+\xi\e{corr}(r_\perp, r_\parallel).
\end{equation}
A final step involves evaluating $\xi\e{model}$ at coordinates $(r,\nu)$ using that $r=\sqrt{r^2_\perp + r^2_\parallel}$ and $\nu=r_\parallel/r$, and computing associated multipole moments as
\begin{equation}
    \xi_\ell^{\rm model}(r) = \frac{2\ell+1}{2}\int_{-1}^{1} \xi\e{model}(r,\nu) \mathcal{L}_\ell(\nu) \dd \nu, 
\end{equation}
where $\mathcal{L}_\ell$ is the Legendre polynomial of order $\ell$.

\begin{figure}
    \centering
    \includegraphics[width=\columnwidth]{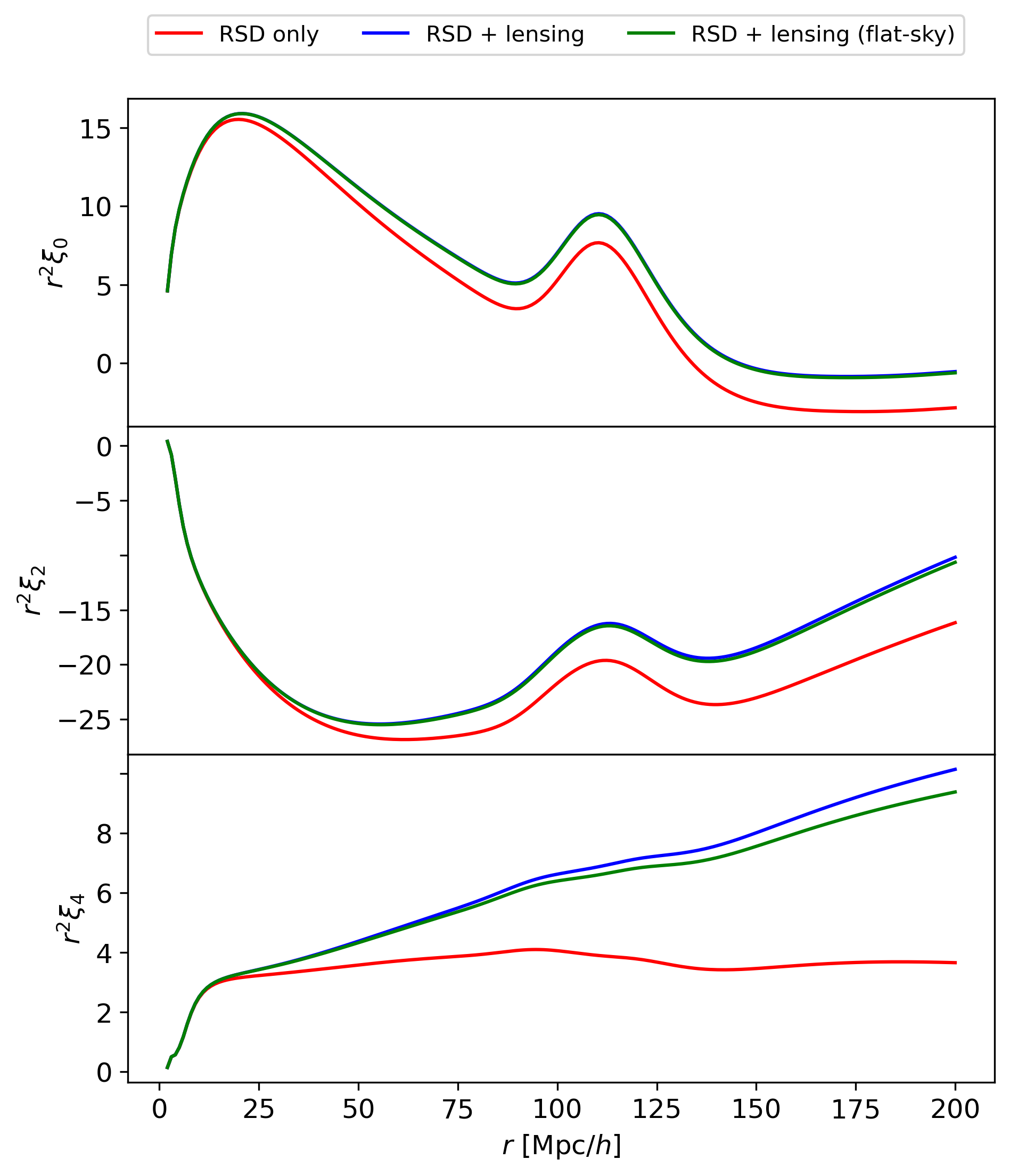}
    \caption{Multipoles of the correlation function (monopole, quadrupole, and hexadecapole from top to bottom) when accounting for RSD only (red), for RSD and lensing magnification (blue), and for RSD and a flat-sky implementation of the lensing magnification. We consider here a $\Lambda$CDM model at $z = 1.8$ with a galaxy bias equal to unity and $s = 1.2$.}
    \label{fig:multipoles_theoretical}
\end{figure}

We show the correlation function multipole moments computed with our model in \cref{fig:multipoles_theoretical}. We considered here the matter at $z=1.8$ in the $\Lambda$CDM model and $s=1.2$. Both flat-sky and curved-sky linear lensing prescriptions are presented. 
We first remark that magnification bias adds a positive contribution to the correlation function multipoles. The case $s = 0$ would also lead to positive contributions for the even multipoles because at this redshift, $|\xi_{\rm len-len}| > |\xi_{\rm den-len}|$, where $|\xi_{\rm len-len}|$ is strictly positive and the sign of $|\xi_{\rm den-len}|$ depends on the value of $s$. While this inequality might not hold at lower redshifts, it is valid for the redshift range we consider, that is, $z > 1$.
Second, the full-sky and flat-sky implementations of the lensing correction give very similar results that are indistinguishable, except on the hexadecapole at large comoving separations. Overall, although we use the curved-sky correction in our modelling, the flat-sky approximation should also work in likelihood analyses, since the covariance associated with the hexadecapole should weakly affect the final results with respect to the dominant monopole and quadrupole.

\section{Methods}
\label{sec:methods}

To investigate the impact of magnification bias on clustering in redshift space, we used $N$-body simulations, which naturally account for the fully non-linear structure formation, and extracted light cones with various magnification bias selections. We then estimated the first three even multipole moments of the two-point correlation function in the light cones in several tomographic redshift bins and ran a Monte Carlo Markov chain (MCMC) likelihood analysis to sample the parameters of the model described in \cref{sec:theory}. We present the different methods in this section.

\subsection{Datasets}
\label{subsec:datasets}

\begin{figure}
    \centering
    \includegraphics[width=\columnwidth]{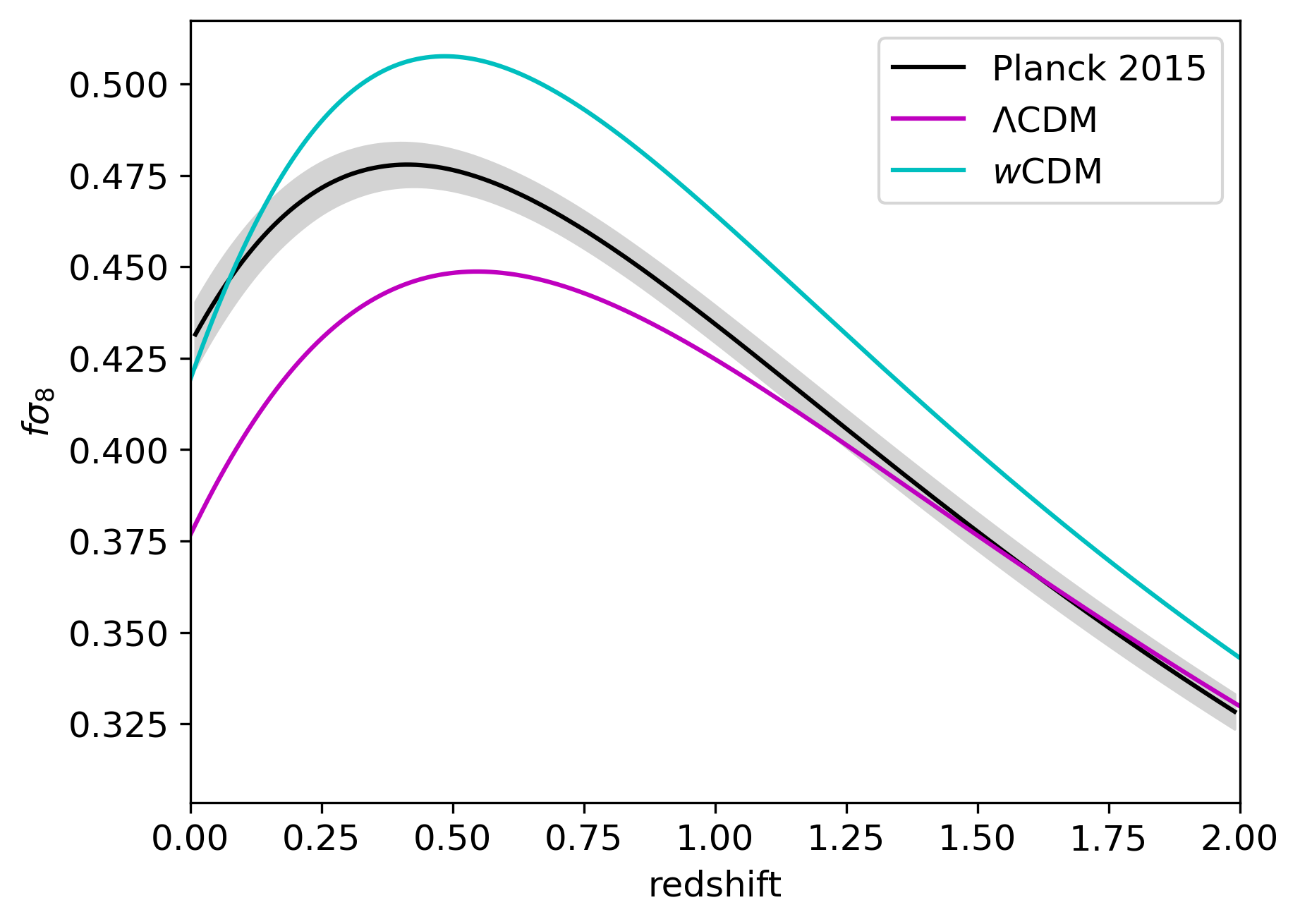}
    \caption{Evolution of the growth rate of structure as a function of redshift for the \cite{planck2016cosmological} cosmology in black (error bars are shown in grey), as well as the two cosmologies ($\Lambda$CDM in purple, $w$CDM in cyan) of the RayGal simulations until $z = 2$.}
    \label{fig:fs8_cosmos}
\end{figure}

The RayGal simulation suite\footnote{\url{https://cosmo.obspm.fr/public-datasets/}} \citep{breton2019imprints,rasera2021raygal} is based on RAMSES \citep{teyssier2002cosmological,guillet2011simple}. These are dark-matter-only $N$-body simulations containing $4096^3$ dark matter (DM) particles of mass $1.8\times10^{10}~M_\odot$ in a volume of $2.625^3~h^{-3}~$Gpc$^3$. Both $\Lambda$CDM and $w$CDM versions are available, and associated fiducial cosmological parameters are given in Table \ref{tab:cosmologicalparameters}. 
The two cosmologies have different $\Omega_m$ and $\sigma_8$ and therefore different values of $f\sigma_8(z)$, since $f \approx \Omega_m(z)^{0.55}$ in General Relativity \citep{wang1998cluster,linder2007parameterized}.%
\begin{table}
        \centering
        \begin{tabular}{cccc} 
                \hline
                \hline
                Model& $\Omega_m$ & $\sigma_8$ & $w$\\
                \hline
                $\Lambda$CDM&0.25733 & 0.80101 & -1.0\\ 
                $w$CDM&0.27508 & 0.85205 & -1.2\\
                \hline \\
        \end{tabular}
        \caption{Cosmological parameters, i.e., $\Omega_m$ the total matter density, $\sigma_8$ the power spectrum normalisation at $z=0$, and $w$ the redshift-independent equation of state for the $\Lambda$CDM and $w$CDM cosmologies of RayGal. In both cases we consider flat models, i.e., $\Omega_k = 0$, with a reduced Hubble parameter $h = 0.72$, a baryon density $\Omega_b$ = 0.04356, a radiation density $\Omega_r = 8\times10^{-5}$ , and a spectral index $n_s = 0.963$.}
        \label{tab:cosmologicalparameters}
\end{table}
This is shown in \cref{fig:fs8_cosmos}, where we show the fiducial values of $f\sigma_8$ as a function of redshift for the two cosmologies as well as the expectations from \citet{planck2016cosmological} $\Lambda$CDM best-fitting model assuming General Relativity.
Interestingly, the values of $f\sigma_8(z)$ for the RayGal $\Lambda$CDM ($w$CDM) model are close those from Planck at high (low) redshift. It is worth emphasising the importance of analysing simulations with different cosmologies, since we can analyse them blindly in a fiducial cosmology, as in observations, and see whether unbiased estimates of the growth rate of structure can be recovered. 

\subsubsection{RayGal light cones}

Several light cones have been extracted from the RayGal simulations. In the present work, we used light cones with an aperture of 2500~deg$^2$ extending to $z = 2$, which encompasses the redshift range probed by the DESI \citep{desi2016desi} and Euclid \citep{laureijs2011euclid} surveys. These light cones contain DM particles as well as DM haloes identified with the parallel friend-of-friend algorithm pFoF \citep{roy2014pfof}, using a linking length of 0.2. We imposed a minimum of 100 particles per halo, which leads to haloes with masses higher than $1.8\times 10^{12}~M_\odot$.

The gravitational-lensing information was computed in the light cones by using the ray-tracing code \textsc{Magrathea-Pathfinder} \citep{reverdy2014propagation,breton2021magrathea}. This code implements an iterative algorithm that finds the null geodesics connecting the observer to each source \citep{breton2019imprints}, that is, either particles or haloes. This allows the computation of RSD and lensing effects at the same time in a general and accurate way. It is important to emphasize the fact that the treatment of gravitational lensing does not involve the Born approximation, which is often used. Our light cones contained about $1.2\times10^7$ haloes in both cosmologies, and we ray-traced about $4\times10^8$ randomly selected particles. Having both haloes and particles enabled us to perform a redshift-space clustering analysis on a biased population for the former (and hence, closer to observations), and for the latter, to carry out a precise study in which the number of matter tracers was maximised.

\begin{figure}
    \centering
    \includegraphics[width=\columnwidth]{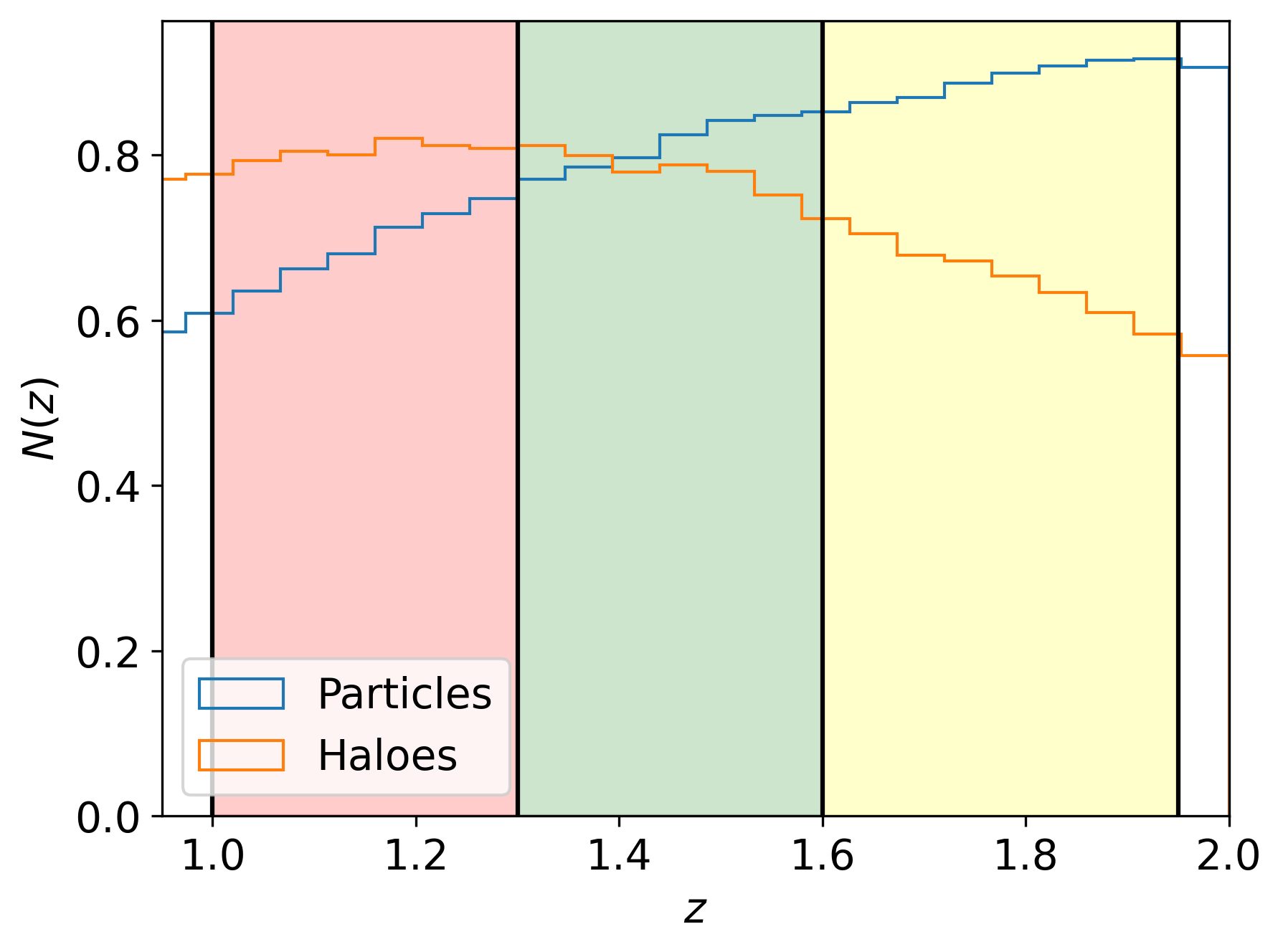}
    \caption{Normalised redshift distribution of the DM particles and haloes, in blue and orange histograms, respectively, within the narrow cone of the $\Lambda$CDM RayGal simulation. The red, green, and yellow regions refer to the redshift bins, i.e. $z = [1.0,1.3], [1.3,1.6],$ and $[1.6,1.95]$.}
    \label{fig:nz_lcdm_raygal}
\end{figure}

In \cref{fig:nz_lcdm_raygal} we show the redshift distribution of the halo and particle samples in the $\Lambda$CDM light cone, as well as the adopted tomographic redshift bins. The distributions in the $w$CDM light cone are very similar.
The tomographic redshift bins cover a similar redshift range as present and future galaxy cosmological surveys, a regime in which gravitational lensing effects on galaxy clustering start to be significant (at about $z>1$). The shape of the redshift distribution increases for particles monotonically, as expected in the case of constant density. However, at about $z = 2$, $N(z)$ seems to decrease. This edge effect arises because we built our light cones up to $z\sim 2$. To avoid any issue, we used a maximum redshift of $z\e{max} = 1.95$. For haloes, $N(z)$ reaches a maximum at around $z = 1.2$ and then decreases. This can be explained by the combined effect of the halo formation and limited mass resolution in the simulation. We did not impose any further selection in redshift to avoid discarding too many objects from our samples, and we thus maximised RSD and lensing magnification signals.

\subsubsection{Implementation of the magnification bias}
\label{subsubsec:magnification_bias_implementation}
To reproduce different levels of magnification bias, a halo mass-galaxy luminosity relation might be imposed and apparent magnified fluxes or magnitudes might be estimated, from which then selections might be made. While it is clearly the appropriate methodology when constructing most realistic mock catalogues, our goal is to investigate the effect of the magnification bias on galaxy clustering in a general way, independent of the properties of any target sample. Hence, we found that the easiest and most efficient way to mimic the effect of magnification bias is to directly use the magnification of sources. Objects may be selected using a probability function proportional to $\mu^{2.5s}$ (see also \cref{subsubsec:lensed_number_counts}). The advantages of this approach is that fewer sources are discarded, and the approach does not depend on the mass resolution of the simulation. However, it depends on some normalisation and discards more sources at high values of $s$. We therefore instead chose to weight each source by $\mu^{2.5s}$. This allowed us to account for the effect of lensing magnification while keeping all the objects in our sample. 
We chose a weight equal to $\mu^{2.5s}$ (in addition to the observed angular positions in the galaxy clustering analysis) because our ray-tracing code computes the distortion matrix along the null geodesics that connect the observer to each source. In this case, the weak-lensing statistics in our sample is not the same as if we were using the Born approximation. Our averaging procedure indeed performs a `source averaging' \citep{kibble2005average,bonvin2015cosmological,kaiser2016bias,breton2021theoretical}. On average, light rays propagate in under-dense regions due to their path on the real null geodesics, which leads to a negative mean convergence. Had we used the Born approximation instead, our full sample would have been `unlensed' and would have led to a vanishing mean convergence. In this case, to implement the effect of magnification in the mock catalogue, we would rather have used the comoving angular positions and applied a weight equal to $\mu^{2.5s-1}$ to each source.

\begin{figure}
    \centering
    \includegraphics[width=\columnwidth]{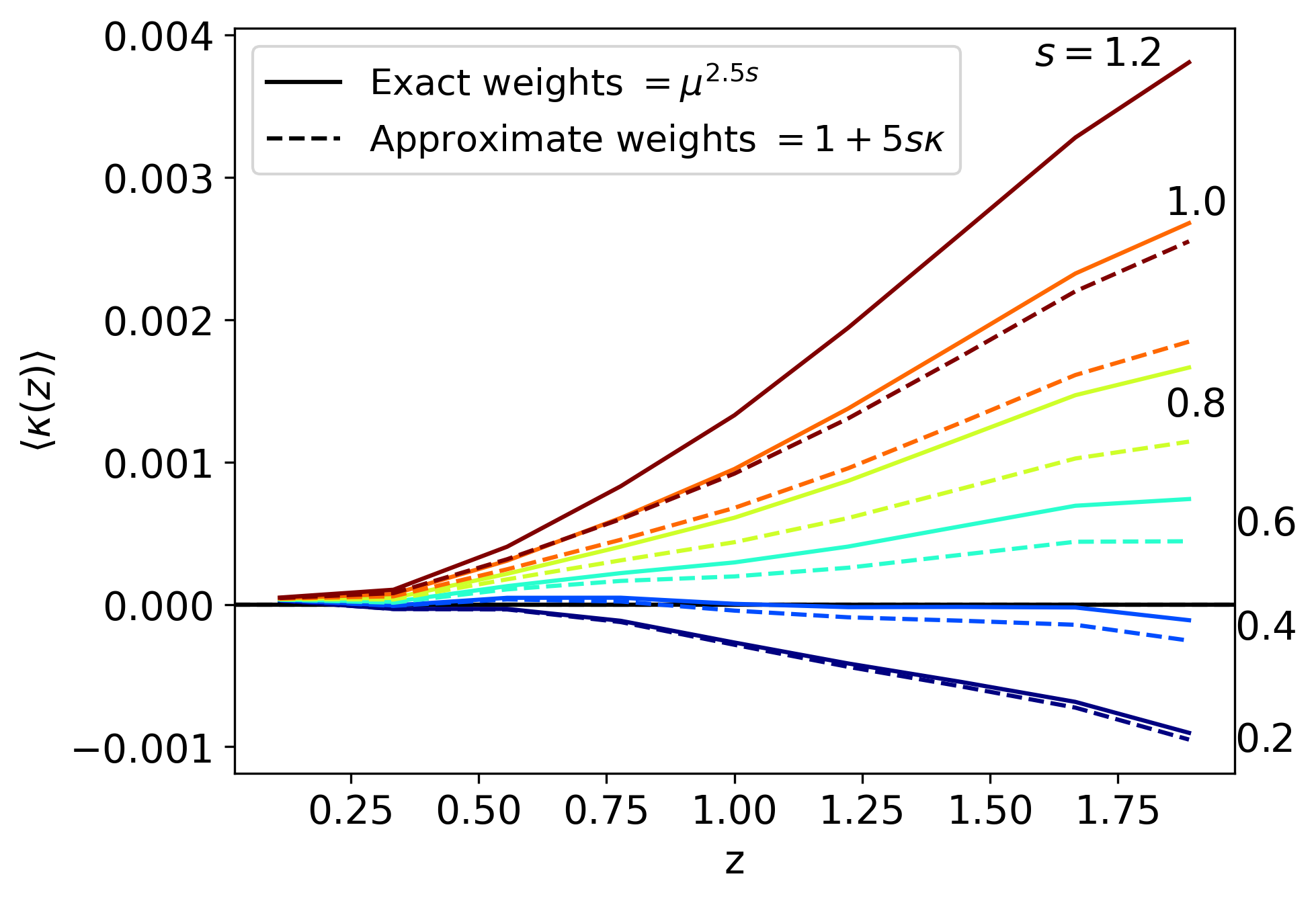}
    \caption{Mean convergence as a function of redshift. The solid and dashed lines refer to the use of weights from the exact and approximate solutions, respectively. We calculated the convergence using the $\Lambda$CDM halo catalogue. The dark blue, blue, cyan, green, orange, and brown lines refer to the mean magnified convergence estimated from an unlensed sample with a factor $s = 0.2, 0.4, 0.6, 0.8, 1.0,$ and $1.2,$ respectively.}
\label{fig:mean_kappa}
\end{figure}

We note that in our theoretical modelling of the magnification bias, we assumed the weak-lensing limit, meaning that the convergence and magnification are small. We therefore used \cref{eq:delta_lens_approx} instead of the exact \cref{eq:delta_lens} for the lensed number count. \cref{fig:mean_kappa} shows the impact of this approximation on the averaged convergence as a function of redshift.
We first note that the mean convergence in the unlensed case (i.e. when $s = 0.4$) is very close to zero, demonstrating the validity of our method to implement the effect of lensing magnification. For $s = 0.2$, the difference between the exact and weak-lensing solution is very small. For higher values, the discrepancy clearly grows. The difference between the exact and weak-lensing solutions increases with $s$, so that for $s = 1.2,$ it reaches $\sim$40\% at $z = 1.9$. The question is whether this is a problem and how it affects the correlation function, since analytical prescriptions only work in the weak-lensing limit. The multipoles of the correlation function in \cref{fig:absdiff_multipoles_lensing} are indeed different for $s = 1.2$, but not by the 40\% of \cref{fig:mean_kappa}. This suggests that the first-order lensing correction should be used with caution in the weak-lensing limit, as its validity depends on redshift and on the value of $s$.


\subsection{Cosmological analysis}
\label{subsec:analysis}

We now turn to the galaxy clustering analysis of the different samples described in \cref{subsec:datasets}. The total number of elements in all the studied cases are summarised in \cref{tab:datasets}.
\begin{table*}
        \centering
        \begin{tabular}{ccccc} 
                \hline
                \hline
                \multirow{2}{*}{Redshift bin} &  \multicolumn{2}{c}{$\Lambda$CDM} & \multicolumn{2}{c}{$w$CDM} \\
                & Particles ($\times 10^6$) & Haloes ($\times 10^6$)& Particles ($\times 10^6$)&  Haloes ($\times 10^6$) \\
                \hline

                $1.0<z<1.3$  &  9.6 & 2.2 & 9.7 & 2.5\\
                       
                $1.3<z<1.6$  &  11.3 & 2.2 & 11.3 & 2.4\\
                      
                $1.6<z<1.95$  &  14.5 & 2.1 & 14.2 & 2.4\\
                  \hline \\
        \end{tabular}
        \caption{Number of objects in the different cases as a function of the redshift bin, type of selection, cosmology, and source type (particles or haloes).}
        \label{tab:datasets}
\end{table*}
Although RayGal simulations provide a full redshift decomposition at first order in metric perturbations, we only focus on the impact of lensing magnification beyond RSD and therefore ignore the more subtle effects that could affect the observed redshift. This means that we only perturb the redshift with the Doppler effect induced by peculiar velocities. 
The unlensed case corresponds to using the comoving angles, while the $s = 0$ case corresponds to using observed angles. For $s > 0$, we took observed angles and weight each source by $\mu^{2.5s}$, as discussed in \cref{subsubsec:magnification_bias_implementation}. In any configuration, the number of elements in \cref{tab:datasets} is the same (modulo tiny differences between observed and comoving angles due to the footprint). For particles, we selected a random sub-sample of the $4\times10^8$ initial particles in our light cones, as the calculation of the anisotropic correlation is very expensive computationally.

\subsubsection{Anisotropic two-point correlation function}


The estimation of three-dimensional clustering necessitates the assumption of a fiducial cosmology to convert angular positions and redshifts into comoving separations. We assumed the $\Lambda$CDM cosmology of RayGal (\cref{tab:cosmologicalparameters}) as our fiducial cosmology, which we used to analyse all the samples in \cref{tab:datasets}, including those extracted from the $w$CDM light cone. The anisotropic correlation function was estimated with 
the Landy-Szalay estimator \citep{landy1993bias} as
\begin{equation}
\label{eq:ls_estimator}
    \xi\e{LS}(r,\nu) = \frac{DD(r,\nu) - 2DR(r,\nu) + RR(r,\nu)}{RR(r,\nu)},
\end{equation}
where $DD$, $DR$, and $RR$ stand for data-data, data-random, and random-random pairs (weighted and normalised by the total number of elements), respectively. We used \textsc{Corrfunc} \citep{sinha2020corrfunc} to count the anisotropic number of pairs in bins of comoving separation $r$ and cosine angle $\nu$. The random samples contain 50 (20) times more objects than haloes (particles). We assigned redshifts in the random catalogues using the shuffling method, which consists of randomly picking redshifts from the data catalogue. Eventually, the multipole moments of the correlation function were obtained from
\begin{equation}
    \xi_\ell(r) = (2\ell+1)\sum_{\nu = 0}^{\nu = 1} \xi(r,\nu)\mathcal{L}_\ell(\nu)\Delta\nu.
\end{equation}
We are only interested in the first non-vanishing even multipole moments of the correlation function: the monopole ($\ell = 0$), the quadrupole ($\ell=2$), and the hexadecapole ($\ell = 4$). We considered a scale range $27.5$ to 127.5$~h^{-1}$Mpc with bins of 5$~h^{-1}$Mpc, and $200$ bins in $\nu$.

\subsubsection{Covariance matrices}

The covariance matrices on single multipole correlation function measurements are estimated analytically assuming Gaussianity as described in \citet{grieb2016gaussian}. Particularly, the covariance matrix between correlation function multipoles $\ell_1$ and $\ell_2$, and between scales $r_1$ and $r_2$, is
\begin{equation}
    C_{\ell_1, \ell_2}(r_1,r_2) = \frac{i^{\ell_1+\ell_2}}{2\pi^2} \int_0^\infty k^2 \sigma^2_{\ell_1, \ell_2}(k) \bar{j}_{\ell_1}(k r_1) \bar{j}_{\ell_2}(k r_2) \dd k,
\end{equation}
where $\bar{j}_{\ell}$ are the bin-averaged spherical Bessel functions and $\sigma^2_{\ell_1, \ell_2}$ are the per-mode covariance multipole moments, both given in \citet{grieb2016gaussian}. The latter function is an integral over the anisotropic power spectrum, which was set here to the corresponding best-fitting RSD model to measurements.

\subsubsection{Likelihood analysis}
\label{subsubsec:likelihood}

We performed a likelihood analysis of the measured monopole, quadrupole, and hexadecapole correlation functions in each sample. The likelihood $\mathcal{L}$ is defined as
\begin{equation}
-2\ln\mathcal{L}(\vec{\vartheta}) = \sum_{i,j}^{N_p}\vec{\Delta}_i(\vec{\vartheta}) C^{-1}_{ij} \vec{\Delta}_j(\vec{\vartheta}),
\end{equation}
where $\vec{\vartheta}$ is the vector of parameters, $\vec{\Delta}$ is the data-model difference vector, $N_p$ is the total number of data points, and $C$ is the covariance matrix. The model we used had six free parameters: $\vartheta=\{f, b_1, b_2, \sigma_v^2, \alpha_\parallel, \alpha_\perp\}$, which correspond to the growth rate, first- and second-order Lagrangian bias parameters, squared velocity dispersion, and two dilation parameters that account for Alcock-Paczynski distortions, respectively. We note that we varied $f$ and not directly $f\sigma_8$ because the model takes as input the linear power spectrum associated with the $\Lambda$CDM simulation at the redshift of interest. We therefore used a fiducial value of $\sigma_8$ to compute the theoretical prediction. We cannot let $f\sigma_8$ free mainly because the value of $\sigma_8$ is degenerate with the growth factor $D_+(z)$. Within linear theory, this is not a problem as $\sigma_8$ can factor out. Within the framework of CLPT, we cannot do this because the non-linear part of the power spectrum is redshift dependent. In any case, although we let $f$ free in the likelihood analysis, we eventually extracted $f\sigma_8$ and compared it to the fiducial value.

We used the dilation parameters along the parallel and transverse direction to account for any change in cosmology with respect to the fiducial one. Formally, these two parameters enter  our formalism as a multiplicative factor on the scales in the anisotropic correlation function: $\xi(\alpha_\perp r_\perp, \alpha_\parallel r_\parallel)$. If the fiducial cosmology is that of the data, we expect $(\alpha_\perp, \alpha_\parallel) = (1,1)$. Otherwise, these are given by the ratios
\begin{eqnarray}
      \alpha_\perp &=& \frac{D_M(z)/r_d}{D_M^{\rm fid}(z)/r_d^{\rm fid}}, \\
      \alpha_\parallel &=& \frac{D_H(z)/r_d}{D_H^{\rm fid}(z)/r_d^{\rm fid}},
\end{eqnarray}
where the superscript `fid' refers to an estimation using the fiducial cosmology, $D_M(z) = (1+z)D_A(z)$ with $D_A$ the angular diameter distance, $D_H(z) = c/H(z)$, and $r_d$ is the sound horizon at the drag epoch. We note that $z$ refers here to the effective redshift of the sample. For the four redshift bins in our case, these are $z = 0.857, 1.151, 1.448, \text{and }1.769$ for haloes and $z = 0.86, 1.155, 1.453,\text{and } 1.777$ for particles.

As argued in \cite{sanchez2020arguments}, using $\sigma_8$ might not be ideal for galaxy clustering analysis, since it relies on a scale given in $h^{-1}$Mpc units that is cosmology dependent. However, as noted in \cite{bautista2021completed}, $\sigma_8$ can still be used as long as we take the changes due to dilation parameters into account. Instead of computing $\sigma_8$ at $8~ h^{-1}$Mpc from a linear power spectrum at $z = 0$, this quantity should be evaluated at $8\alpha~ h^{-1}$Mpc, where $\alpha = \alpha_\perp^{2/3}\alpha_\parallel^{1/3}$ is the isotropic dilation parameter. In the following, our estimates of $f\sigma_8$ implicitly account for this correction, while the fiducial values are those of \cref{fig:fs8_cosmos}.

We performed three types of analysis. First, a standard analysis that only accounted for RSD using the CLPT-GS model. This is similar to what is routinely done in observational studies of galaxy clustering. Second, an analysis that included the magnification bias correction in the model and where $s$ was fixed to its fiducial value. Although it is possible to estimate $s$ from the data itself, the standard method consists of fitting the local slope at the faint-end of the luminosity function, which might not be accurate when the selection function is complex \citep{vonwietersheim-kramsta2021magnification}. We therefore also propose an alternative analysis that includes the magnification bias correction in the model, but allows the $s$ parameter to vary. Finally, we used the flat priors displayed in \cref{tab:priors} (where only the last analysis allows $s$ to vary).
\begin{table}
        \centering
        \begin{tabular}{cc} 
                \hline
                \hline
                Parameter & Flat prior \\
                \hline
                $f$ & [0, 2]\\
                $b_1$ & [-0.5, 3]\\
                $b_2$ & [-70, 70]\\
                $\sigma_v^2$ & [0, 100]\\
                $\alpha_\perp$ & [0.5, 1.5]\\
                $\alpha_\parallel$ & [0.5, 1.5]\\
                $s$ & [0, 2.5]\\
                \hline \\
        \end{tabular}
        \caption{Flat priors used in the MCMC likelihood analyses.}
        \label{tab:priors}
\end{table}
%
For each sample, we produced MCMCs using the \textsc{Emcee} algorithm \citep{foreman-mackey2013emcee} 
with 100 walkers and 50,000 steps per walker. The walkers were initialised around the fiducial values (and a first-order Lagrangian bias of 0 and 1.4 for particles and haloes, respectively), to which we added a random shift using a Gaussian distribution with zero mean and variance unity, multiplied by 0.1. 
The burn-in phase depended on the auto-correlation time, and chains were thinned by removing highly correlated steps (the thinning was mostly used for convenience when making the plots, as we checked that it did not change the final results). Furthermore, the size of every chain was longer than 50 times the auto-correlation length.
We also verified that we were able to recover the same results for the analysis of DM particles using MultiNest\footnote{We ran MultiNest with 5,000 live points, \texttt{evidence\_tolerance=0.3,} and \texttt{sampling\_effiency=parameter.}} \citep{feroz2008multimodal,feroz2009multinest,feroz2019importance}.
The chains were finally analysed with \textsc{GetDist} \citep{lewis2019getdist} to obtain the final parameter constraints.

\section{Results}
\label{sec:results}

In this section, we present the results of the redshift-space clustering analysis with and without magnification bias for the particle and halo samples. We first discuss the recovered value $f\sigma_8$ and its bias with respect to the fiducial value. We then turn to an overview of the results for the other five (or six when $s$ is free) parameters of the model.

\subsection{Bias on $f\sigma_8$}
\label{subsec:bias_fs8}


\subsubsection{Particles}

\begin{figure*}
    \centering
    \includegraphics[width=2\columnwidth]{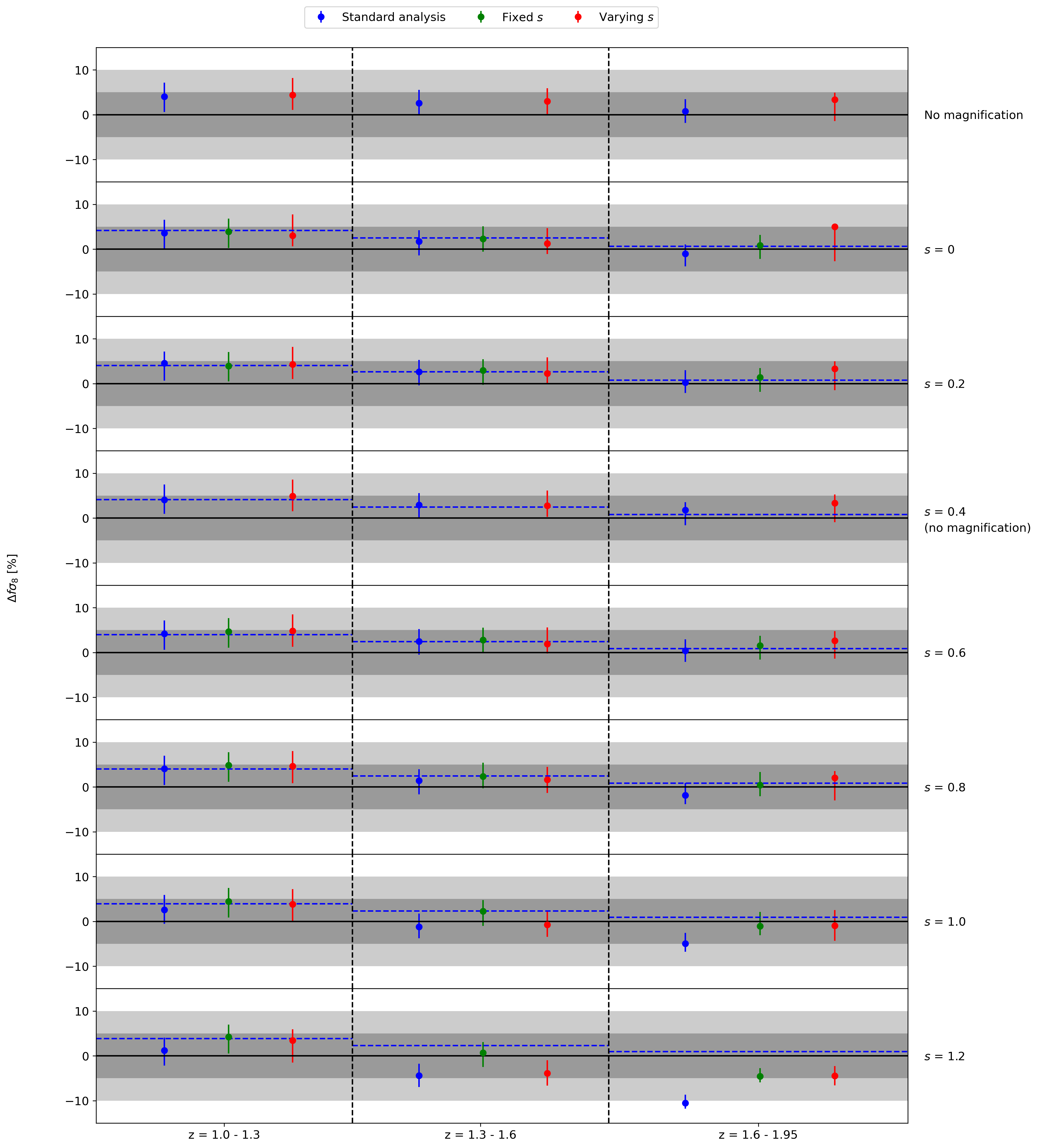}
    \caption{Relative difference on $f\sigma_8$ with respect to its fiducial value as a function of redshift bins and sample configuration for $\Lambda$CDM particles. The data points in blue, green, and red refer to different analyses where we only account for RSD, add magnification bias with $s$ fixed and \textit{s} free, respectively (see also \cref{subsubsec:likelihood}). The central values are those of the overall best fit, while error bars are those of the 68\% confidence interval on the marginalised distribution. The dashed blue lines show the value for the unmagnified case. The grey shades highlight the $\pm$ 5\% and 10\% ranges.}
    \label{fig:fs8_lcdm_part}
\end{figure*}

We first consider the case in which sources are DM particles, as these should give the clearest trend on the impact of magnification bias. The full results for the $\Lambda$CDM case are shown in \cref{fig:fs8_lcdm_part}.
%

In the standard analysis (RSD only), the top row for the unlensed sample shows that we recover the expected value of $f\sigma_8$ within or close to 1$\sigma$ statistical  uncertainty. When lensing magnification is incorporated in the data, it clearly has almost no impact in the first redshift bin at $z = 1.0-1.3$, but strongly shifts the estimate of $f\sigma_8$ at higher redshifts. More precisely, magnification bias leads to an underestimation of the growth rate if it is not modelled. The shift on $f\sigma_8$ does not strictly increase with $s$, which makes sense as the most dominant term in the lensing correction is $\xi\e{lens-lens}$ , which scales as $(5s-2)^2$ (see \cref{subsec:magnification_bias}). This means that $s = 0$ should lead to a similar effect as $s = 0.8$, while $s = 0.4$ completely cancels the lensing effect. Then, the shift on the growth rate monotonically increases for $s > 0.8$. At $z = 1.3-1.6$ and $z = 1.6-1.95$, the shift on $f\sigma_8$ for $s = 1.2$ reaches approximately 9\% and 11\%, respectively, with respect to the unlensed case, while it is 5\% and 6\% for $s = 1$. In the highest redshift bin, we find that the  growth rate is not recovered within $1\sigma$ statistical uncertainty for $s \gtrsim 1$.  

\begin{figure}
    \centering
    \includegraphics[width=\columnwidth]{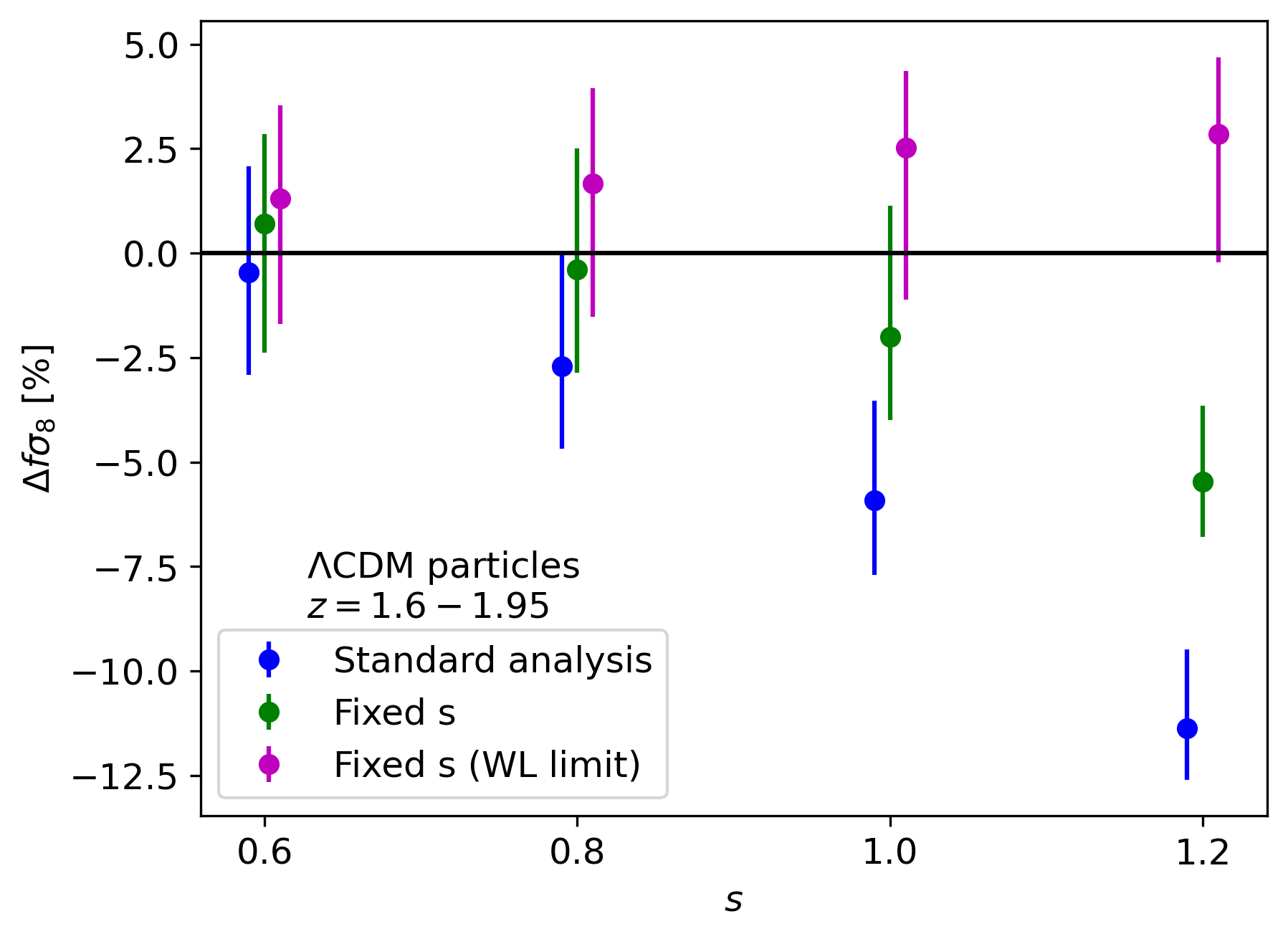}
    \caption{Relative difference on $f\sigma_8$ with respect to the `no magnification' case of \cref{fig:fs8_lcdm_part}. The results for the standard analysis and fixed s of \cref{fig:fs8_lcdm_part} are shown in blue and green, and the results of a fixed s analysis are shown in purple.\ For this last analysis, we implemented magnification in data under the weak-lensing limit, i.e. we  used a weight equal to $\mu(1+(5s-2)\kappa)$ instead of $\mu^{2.5s}$ (see also \cref{subsubsec:magnification_bias_implementation}).}
    \label{fig:reldiff_fs8_lcdm_part}
\end{figure}

When magnification bias is accounted for in the modelling and $s$ is known, we obtain an unbiased estimate of the growth rate with respect to the unmagnified case, except for $s = 1.2$ in the highest redshift bin. This shows that our linear-theory-based lensing correction is good enough overall to correct for the effect of magnification bias. However, for very high values of $s$ at high redshift, the weak-lensing limit that was used to compute the correction might no longer hold (see also the difference in the quadrupole in \cref{fig:absdiff_multipoles_lensing}, where the theoretical prediction seems to underestimate the signal). Therefore, accounting for higher-order terms may be important in the future for the most extreme configurations. In \cref{fig:reldiff_fs8_lcdm_part} we show the relative difference on $f\sigma_8$ with respect to the unmagnified case when the magnification implementation in the data is done in the weak-lensing limit.
As described in \cref{subsubsec:lensed_number_counts}, and \cref{subsubsec:magnification_bias_implementation}, to account for this approximation, we need to use a weight equal to  $\mu(1+(5s-2)\kappa)$ instead of $\mu^{2.5s}$, the prefactor $\mu$ is used to correct for the dilation effect already present in the data. Within the weak-lensing limit (in purple), the relative difference increases only weakly with $s$ and reaches a lower value at $s = 1.2$ (compared with the other cases), and it remains within 1$\sigma$ with respect to the expected value. The difference between the green and purple data points only comes from the weak-lensing limit, and it reaches roughly $8\%$ at $s=1.2$. This shows that it is important to consider higher-order modelling for high-redshift samples with high $s$ values, beyond the first-order weak-lensing limit.

Lastly, when $s$ is free in the fit, the error bars on the parameters are larger, which is expected since there is more freedom in the model. Letting $s$ free might not be the best strategy in the full-shape RSD analysis. Estimated values of $s$ appear to be better suited. Nonetheless, the best-fit values obtained when letting $s$ free are very close to the case where $s$ is fixed, meaning that we are able to estimate this nuisance parameter without any strong bias.

Overall, we find that depending on the sample selection, galaxy clustering surveys probing redshifts above $z\sim 1.3$ will need to properly include lensing magnification as part of the theoretical model to recover unbiased estimates of the growth rate of structure. This will in turn enable them to test gravity while keeping theoretical systematic errors under control.

\begin{figure*}
    \centering
    \includegraphics[width=2\columnwidth]{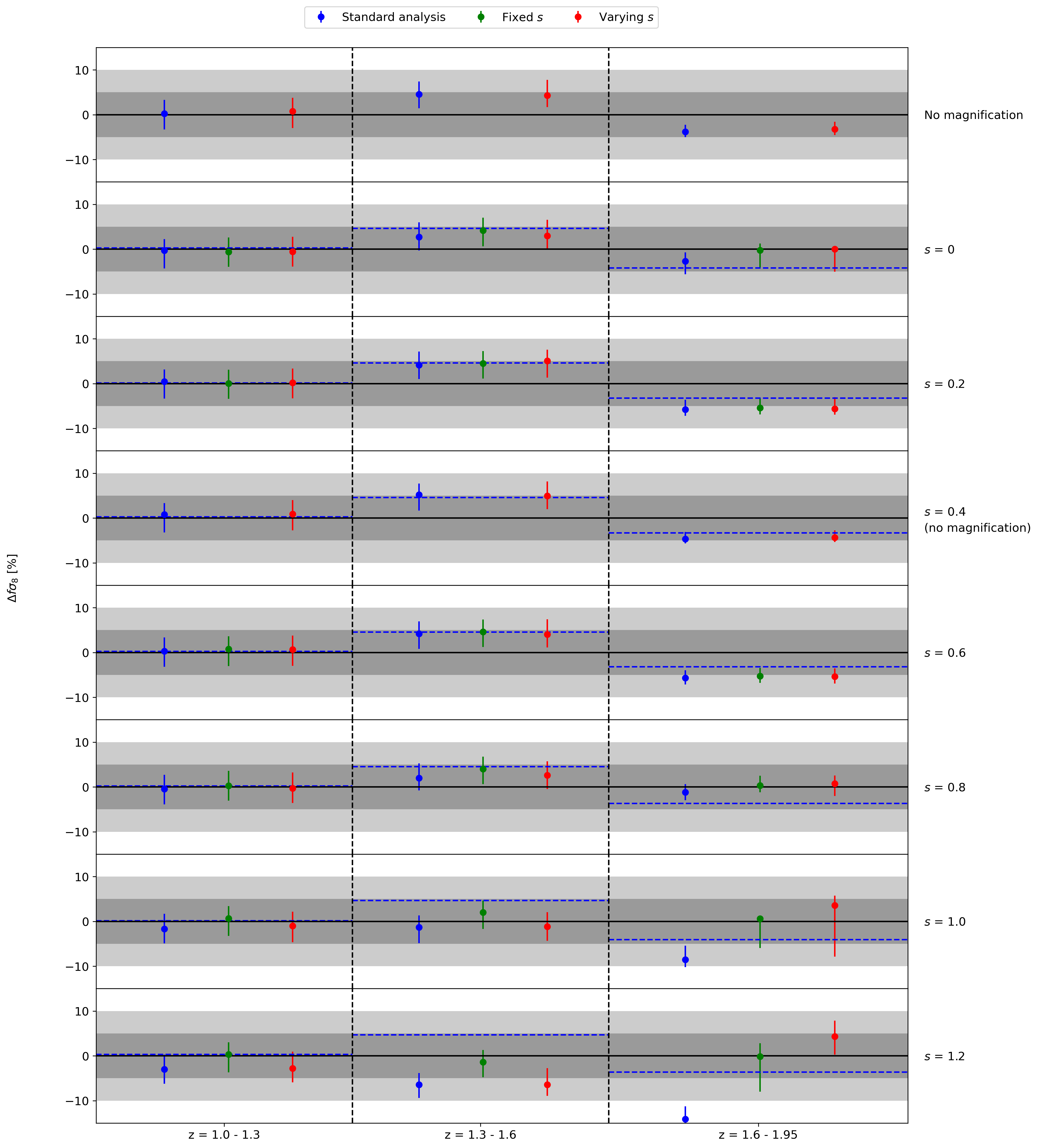}
    \caption{Same as \cref{fig:fs8_lcdm_part}, but for $w$CDM.}
    \label{fig:fs8_wcdm_part}
\end{figure*}

In \cref{fig:fs8_wcdm_part} we show the same analysis, but for the $w$CDM model. 
The results are extremely similar to those of \cref{fig:fs8_lcdm_part}, which is reassuring since here we compare the estimated $f\sigma_8$ with the fiducial values in the $w$CDM cosmology. This means that, although the full analysis is performed using a $\Lambda$CDM fiducial model, we are still able to recover almost unbiased constraints on the cosmology of the data (see also \cref{fig:fs8_cosmos} for the difference in $f\sigma_8$ between the two cosmological models). The error bars for $s = 0.2$ and $s=0.6$ in the highest redshift bin are very large. The fact that these configurations give similar results is expected since the leading magnification bias term scales as $(5s-2)^2$. Furthermore, it seems that for these particular configurations, there is a multi-modal behaviour in the joint posterior probability for the growth rate and second-order Lagrangian bias parameters, leading to larger error bars.
In the last two redshift bins, the shift on $f\sigma_8$ reaches approximately 12\% and 10\% for $s = 1.2$ when the magnification bias is not accounted for. As previously, knowing the true value of $s$ in the modelling gives an unbiased estimate of $f\sigma_8$, except for the last redshift bin for $s = 0.8,$ which seems slightly overestimated.
This shows that although the lensing correction computed with the $\Lambda$CDM model is not exact for the $w$CDM simulation as seen in \cref{appendix:magnification_bias_multipoles}, it still includes most of the effect. However, there can be large discrepancies at high redshift for high values of $s$ for the lensing contribution with respect to the $\Lambda$CDM model (see \cref{fig:absdiff_multipoles_lensing}), and it is therefore not clear that using a correction from the fiducial model is enough when the cosmologies are very different. Even in this case, however, it seems that in our data, adding the magnification bias correction at fixed $s$ gives a much better agreement to the fiducial value of $f\sigma_8$. In the future, it will thus be important to properly model the magnification bias term,
as it could lead to additional bias.

\subsubsection{Haloes}

\begin{figure*}
    \centering
    \includegraphics[width=2\columnwidth]{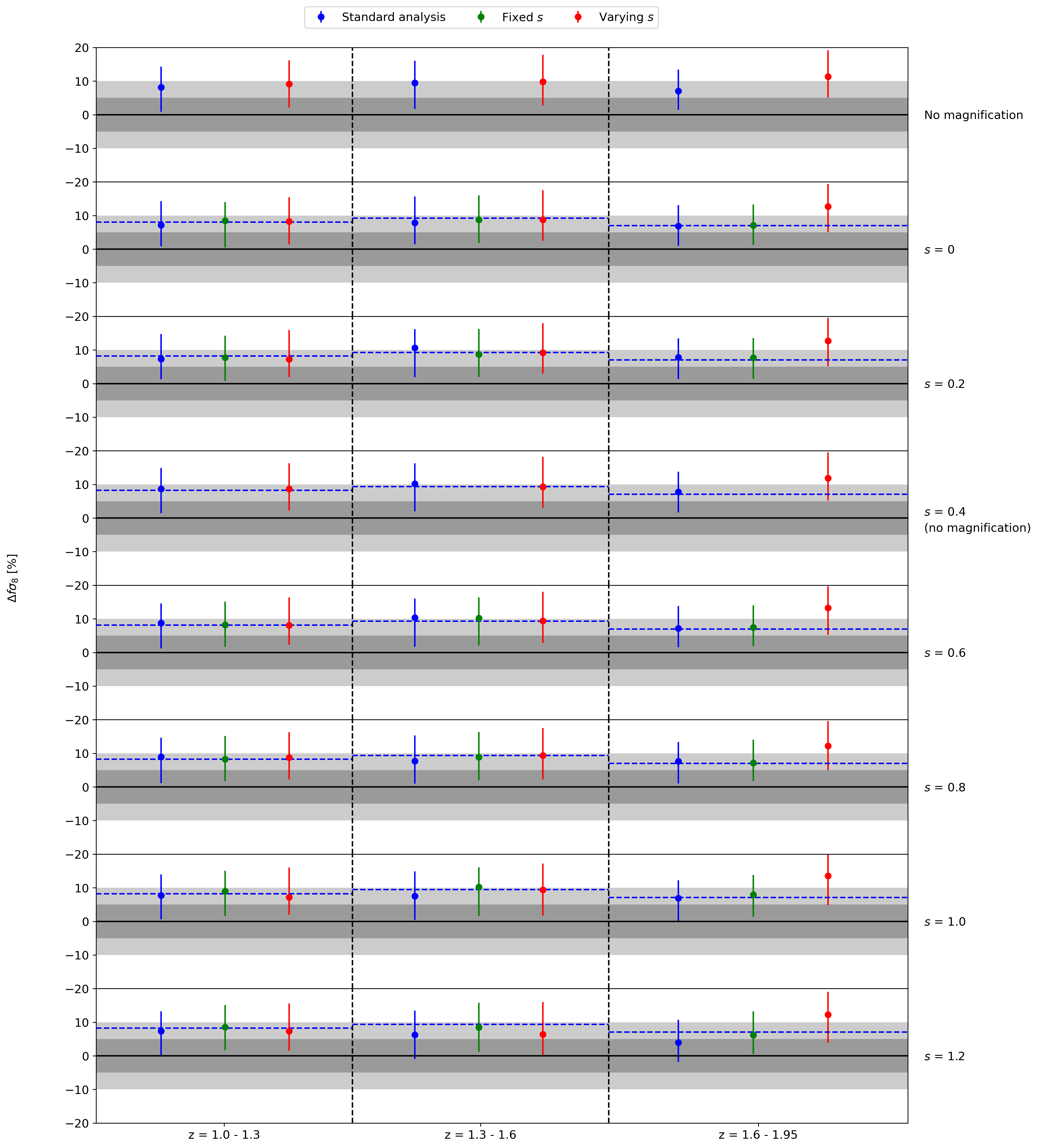}
    \caption{Same as \cref{fig:fs8_lcdm_part}, but for haloes.}
    \label{fig:fs8_lcdm_halo}
\end{figure*}

We now turn to the analysis of DM haloes. While DM particles are useful since they allow characterising the effect of magnification bias with great precision, analysing haloes is more observationally relevant since, just like galaxies, they are biased tracers of the matter density field. This is particularly important because the relative effect of magnification bias on the total number counts scales as $(5s-2)/b$, where $b$ is the linear Eulerian galaxy bias. This means that we expect lensing to have less impact on samples with large bias. In our case, the best-fit RSD-only models for the unmagnified halo samples in the $\Lambda$CDM cosmology gives roughly $b = 2.1, 2.5,$ and 3 in the different redshift bins, from low to high redshifts. We show the results in \cref{fig:fs8_lcdm_halo}.
Qualitatively, the results are similar to those of \cref{fig:fs8_lcdm_part}. We find that the values of the growth rate tend to be slightly overestimated, even when the sample is not magnified. This can be attributed to residual theoretical uncertainties of the RSD model and to the modest volume probed by our light cones, which can lead to non-negligible sample variance effects. Nonetheless, we can study and discuss the relative differences in the estimated parameters for the various cases. As previously, we find that the central values of $f\sigma_8$ are shifted towards lower values when lensing is not accounted for in the modelling. The main difference between DM particles and haloes is that for the latter, the effect of lensing is not clear as the magnified and unmagnified cases agree within 1$\sigma$ error. In the highest redshift bin, the best-fit value for $s = 1.2$ is shifted by roughly $3\%$ with respect to the unmagnified case. While it is still large, it does not reach the $11\%$ seen for the particles. 

\begin{figure*}
    \centering
    \includegraphics[width=2\columnwidth]{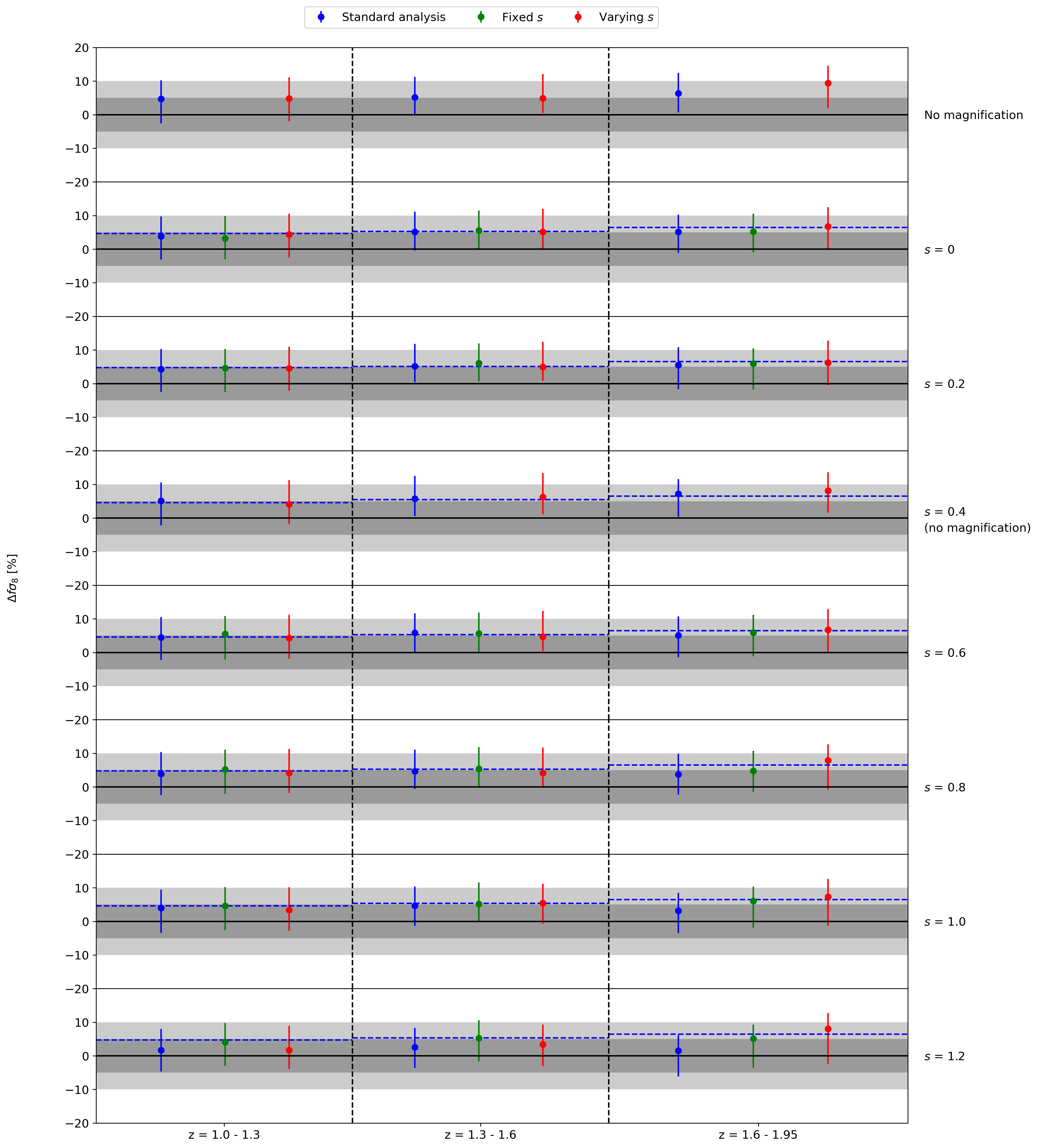}
    \caption{Same as \cref{fig:fs8_lcdm_halo}, but for $w$CDM.}
    \label{fig:fs8_wcdm_halo}
\end{figure*}

The results displayed in \cref{fig:fs8_wcdm_halo} for $w$CDM haloes are very similar.
We find a discrepancy of $3.5\%$ and $5.5\%$ in the highest redshift bin for $s = 1.0, 1.2$ with respect to the unmagnified case. While larger than for $\Lambda$CDM haloes, it still does not reach the $10\%$ difference that we have for particles. 

Overall, we find that the effect of magnification bias is lower for haloes than for particles, which is expected due to galaxy bias. Moreover, the relative difference on $f\sigma_8$ is suppressed by a factor close to $b$. This means that very biased target samples might not be significantly subject to lensing effects at the considered redshifts in the present work.

\subsection{Other parameters}
\label{subsec:results_other_parameters}

\begin{figure*}
    \centering
    \includegraphics[width=2\columnwidth]{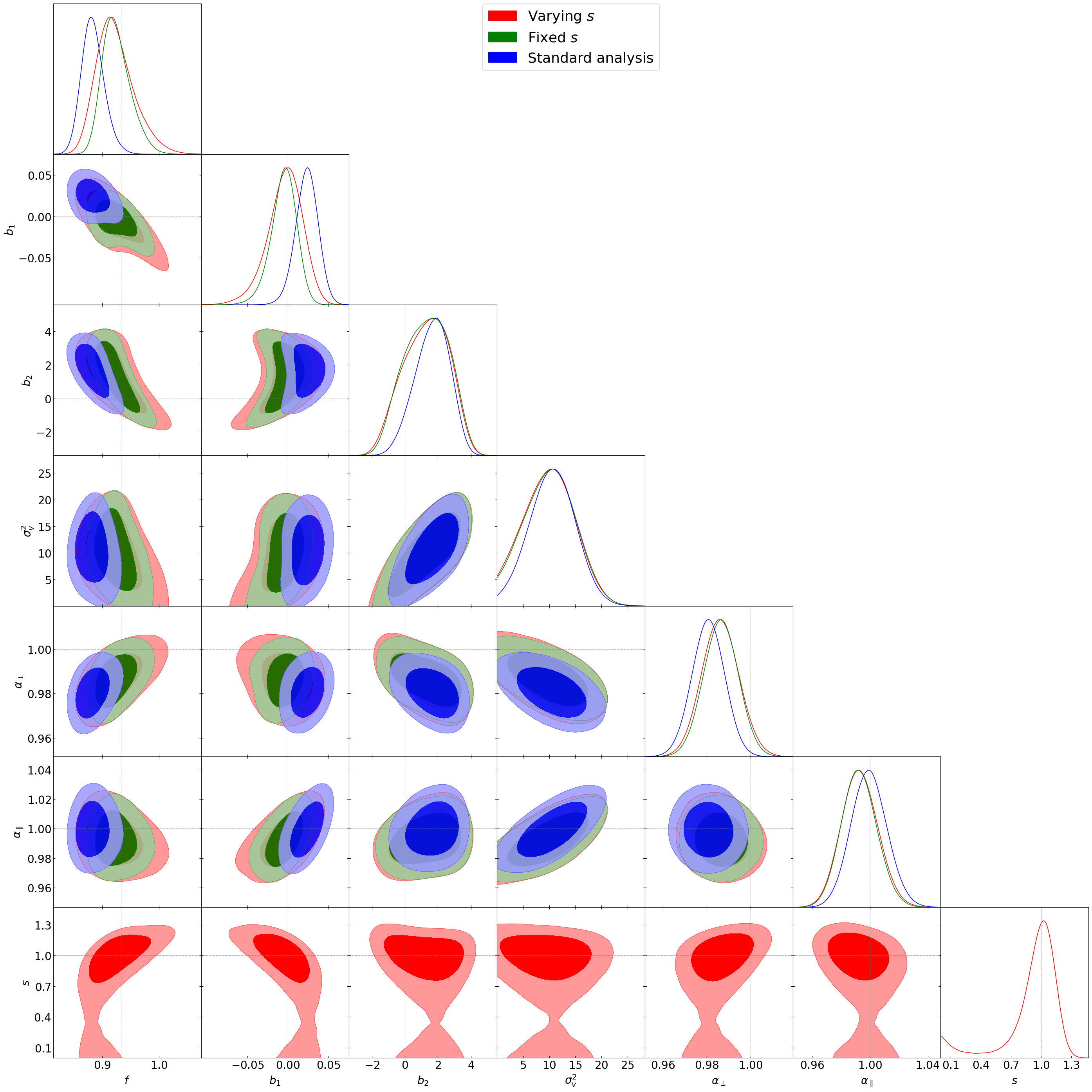}
    \caption{Posterior distributions for the parameters of our model for the $\Lambda$CDM particle sample with $s = 1.0$ at $z = 1.6-1.95$ (see also the full results in \cref{fig:fs8_lcdm_part}).}
    \label{fig:triplot_lcdm_s1p0}
\end{figure*}

In this section we study the impact of magnification bias on the parameters of our model beyond the growth rate of structure. We consider $\Lambda$CDM particles in two cases: $s = 1.0$, as it is a realistic case for future spectroscopic surveys \citep{lepori2021impact}, and $s = 1.2$  where the effect of magnification bias is the most significant. We present the posterior distributions for $s = 1.0$ in \cref{fig:triplot_lcdm_s1p0}.
%
When lensing is not accounted for, the estimate of the growth rate is biased, while the lensing correction (when $s$ is fixed or free) allows us to recover the fiducial value within an error of 1$\sigma$ . These results are approximately the same as those of \cref{fig:fs8_lcdm_part}, except that we show $f$ (the free parameter of our model) and not $f\sigma_8$. Since our fiducial cosmology is already that of the data, there should not be any qualitative difference, in the sense that we only need to multiply $f$ by the fiducial $\sigma_8$ (because the dilation parameters should be very close to unity in principle).
For DM particles, the first-order Lagrangian bias $b_1$ should be equal to zero, and this is indeed what we obtain. However, not accounting for lensing magnification leads to an overestimation of the bias. This comes from the fact that lensing adds a positive contribution to the multipole moments (as shown in \cref{fig:multipoles_theoretical}), which is counter-balanced in the likelihood analyses by higher values of the galaxy bias. Moreover, the $(f,b_1)$ joint posterior probability clearly shows an anti-correlation, where higher values of the linear bias induce lower values of $f$. This comes from the fact that higher values of the galaxy bias lead to a larger amplitude of the quadrupole. In order to fit the data, the likelihood analysis therefore converges towards a lower value of $f$. 

For the growth rate and galaxy bias parameters, the error bars increase when \textit{s} is let free with respect to the case when $s$ is fixed. However, this is not the case for the other nuisance parameters, that is, $b_2$, $\sigma_v^2$ , and dilation parameters, where the two analyses give similar marginal distributions. Regarding the estimation of $s$, we see a remarkable agreement between the expected and best-fit values. This shows that, in principle, we should be able to accurately recover $s$ through a galaxy clustering analysis. Furthermore, we notice that the marginal distribution of $s$ exhibits a multi-modal behaviour around $s = 0$.  This comes from the fact that the leading correction term is proportional to $(5s-2)^2$, and there are two plausible possibilities when $s \lesssim 0.8$. This is a further evidence that it might not be optimal to let $s$ free with a flat uninformative prior.

Finally, we find that in none of the cases, $(\alpha_\perp, \alpha_\parallel) = (1,1)$ is recovered exactly within an error of 1$\sigma$ error, although our results are very close. This might be related to the fact that we have one particular realisation of the density field in a limited volume, which is subject to sample variance. We note that in the RSD-only case, the estimated $\alpha_\perp$ and $\alpha_\parallel$ parameters depart from fiducial values more than when lensing is accounted for. This means that in this case, the likelihood analysis prefers other cosmological models. This might further impact the estimation of the dilation-corrected $\sigma_8$ , as described in \cref{subsubsec:likelihood}. However, this additional bias seems low for the redshift bins that we considered. In the case of the second-order Lagrangian bias $b_2$ , the fiducial value is only recovered when lensing is accounted for, while the inferred values of the velocity dispersion parameter $\sigma_v^2$ are very similar in all cases. Nonetheless, these two parameters are considered as nuisance parameters over which we marginalised.

\begin{figure*}
    \centering
    \includegraphics[width=2\columnwidth]{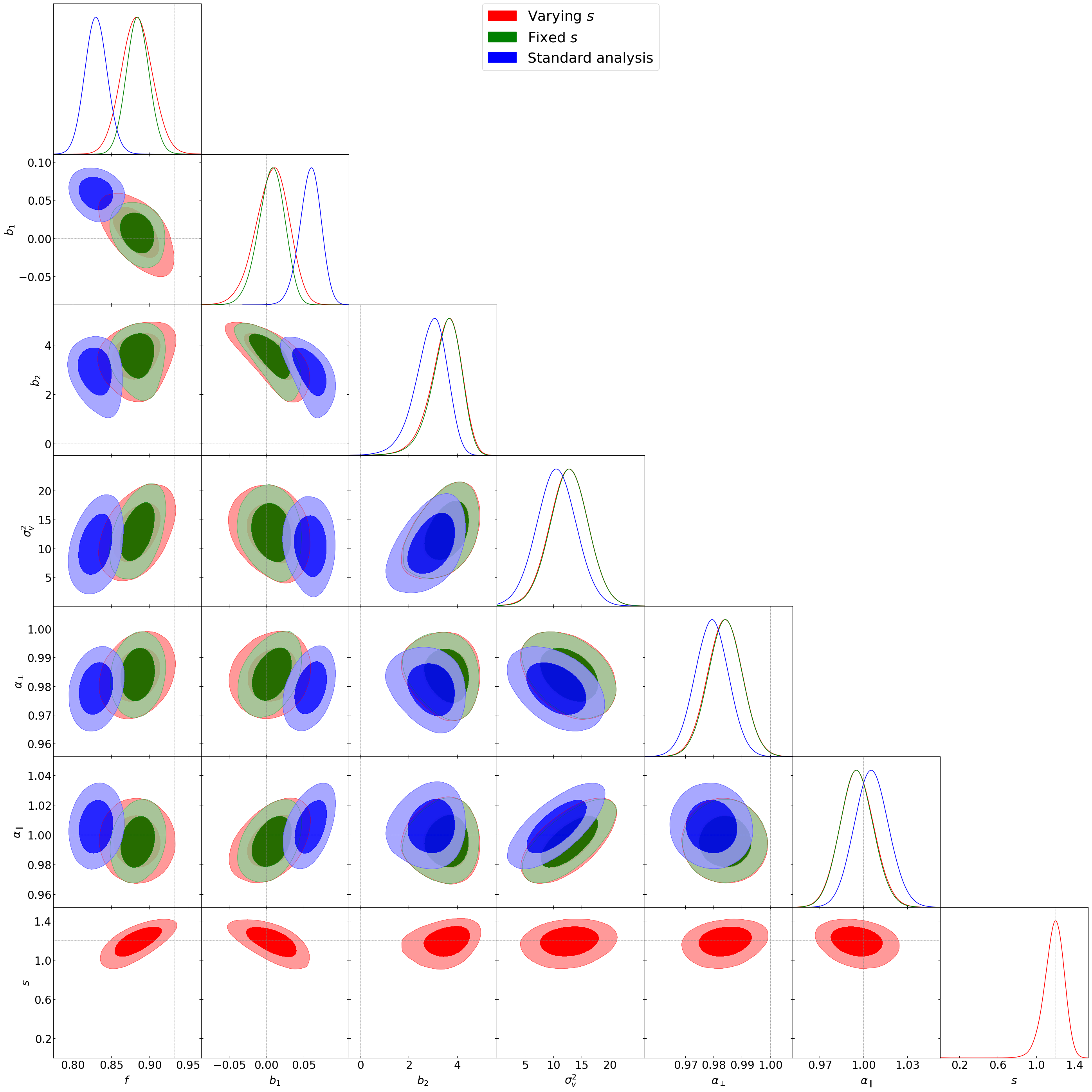}
    \caption{Same as \cref{fig:triplot_lcdm_s1p0}, but with $s = 1.2$.}
    \label{fig:triplot_lcdm_s1p2}
\end{figure*}

Lastly, we present the posterior distributions for $s = 1.2$ in \cref{fig:triplot_lcdm_s1p2}.
The results are similar to those of \cref{fig:triplot_lcdm_s1p0}, except that here we do not recover the value of $f$ within an error of 1$\sigma$ , even when we account for magnification bias as seen in \cref{fig:fs8_lcdm_part}. As in \cref{fig:triplot_lcdm_s1p0}, we remark that we are able to recover $s$ accurately, even in this configuration, in which the weak-lensing limit might start to hold no longer. We also note that there is no multi-modal behaviour, in contrast to the case with $s = 1.0$. This is because $s$ is large enough so that there is no ambiguity between the different solutions at $s < 0.8$. This shows that, generally, adding the lensing contribution in the modelling allows a considerably better recovery of the fiducial parameters.

\section{Conclusion}
\label{sec:conclusion}

We have studied the impact of lensing magnification on the three-dimensional galaxy clustering in redshift space and how the growth rate of structure is affected when lensing magnification is not properly accounted for in the modelling. This effect, commonly referred to as magnification bias, comes from the fact that gravitational lensing modifies the apparent angular positions of sources and magnifies their fluxes. Therefore, magnitude-limited surveys are sensitive to it as it generates additional apparent spatial correlations in the data.
We performed an exhaustive galaxy clustering analysis of the multipole moments of the correlation function, where the theoretical model relied on the convolution Lagrangian perturbation theory and Gaussian streaming model for non-linear RSD and on a curved-sky, linear-theory-based lensing correction. We compared this model to high-resolution $N$-body simulations with two different cosmologies, $\Lambda$CDM and $w$CDM, where the effects of RSD and magnification bias were implemented in simulated catalogues of DM particles and haloes.


The main results of our paper are the following. First, magnification bias only becomes relevant at $z > 1.3$. Second, not accounting for magnification bias in the modelling gives a biased estimate of the growth rate. We found that analysing a magnified sample with a theoretical model that only contains RSD leads to an overestimation of the galaxy bias, and more importantly, to an underestimation of the growth rate. Depending on the redshift and slope of the galaxy luminosity function, the best-fit value of $f\sigma_8$ can be shifted by more than 10\%.
It appears that using the linear-theory lensing correction allows the recovery of unbiased estimates of the growth rate in most cases when the lensing contribution is small if $s$ is accurately known a priori. We also showed that for the cases with high redshift, large $s,$ and low bias, it could be important to model the lensing contribution beyond the first-order weak-lensing limit.
If $s$ is unknown, it might be tempting to keep it as a free parameter of the model. While this allows the recovery of an unbiased estimate of $s$, it significantly increases the uncertainty on $f\sigma_8$, which we might refrain from. This shows that $s$ should not be let free (or at least, no flat uninformative prior should be used on that parameter) and a way should be found instead to estimate it differently.
Finally, we find that we were able to recover the same results for a $w$CDM simulation, although it was analysed with a fiducial $\Lambda$CDM model. This is encouraging and shows that in our case, the modelling is reliable even if we do not know the underlying cosmology of the data exactly. However, further investigations might be needed to estimate the impact of the cosmological model on the lensing correction as it could in principle be another source of bias.

Gravitational lensing will play an important role in galaxy clustering analyses in future high-redshift surveys, and its effect will have to be implemented in order to accurately recover cosmological parameters. This effect has not been considered in observational studies so far, even when the redshift of the sample was high \citep{okamura2016subaru,zarrouk2018clustering,hou2021completed}. However, this does not necessarily mean that these works were not correct because statistical uncertainties are still significant and  the effect of the magnification bias depends on the properties of the galaxy sample under scrutiny.
An interesting prospect for future works would be to consistently incorporate the effect of gravitational lensing within the theoretical framework of non-linear RSD. Nonetheless, we showed that in our case, a hybrid model of non-linear RSD and linear lensing is in most cases sufficient for the accuracy of our simulated data.

\begin{acknowledgements}
This work was granted access to HPC resources of TGCC/CINES through allocations made by
GENCI (Grand Équipement National de Calcul Intensif) under the allocation 2020-A0070402287.
The project leading to this publication has received funding from Excellence Initiative of Aix-Marseille University - A*MIDEX, a French “Investissements d'Avenir” programme (AMX-19-IET-008 - IPhU).
\end{acknowledgements}

\bibliographystyle{aa}
\bibliography{biblio} 

\appendix

\section{Magnification bias modelling of the correlation function multipoles} \label{appendix:magnification_bias_multipoles}

\begin{figure*}
    \centering
    \includegraphics[width=\columnwidth]{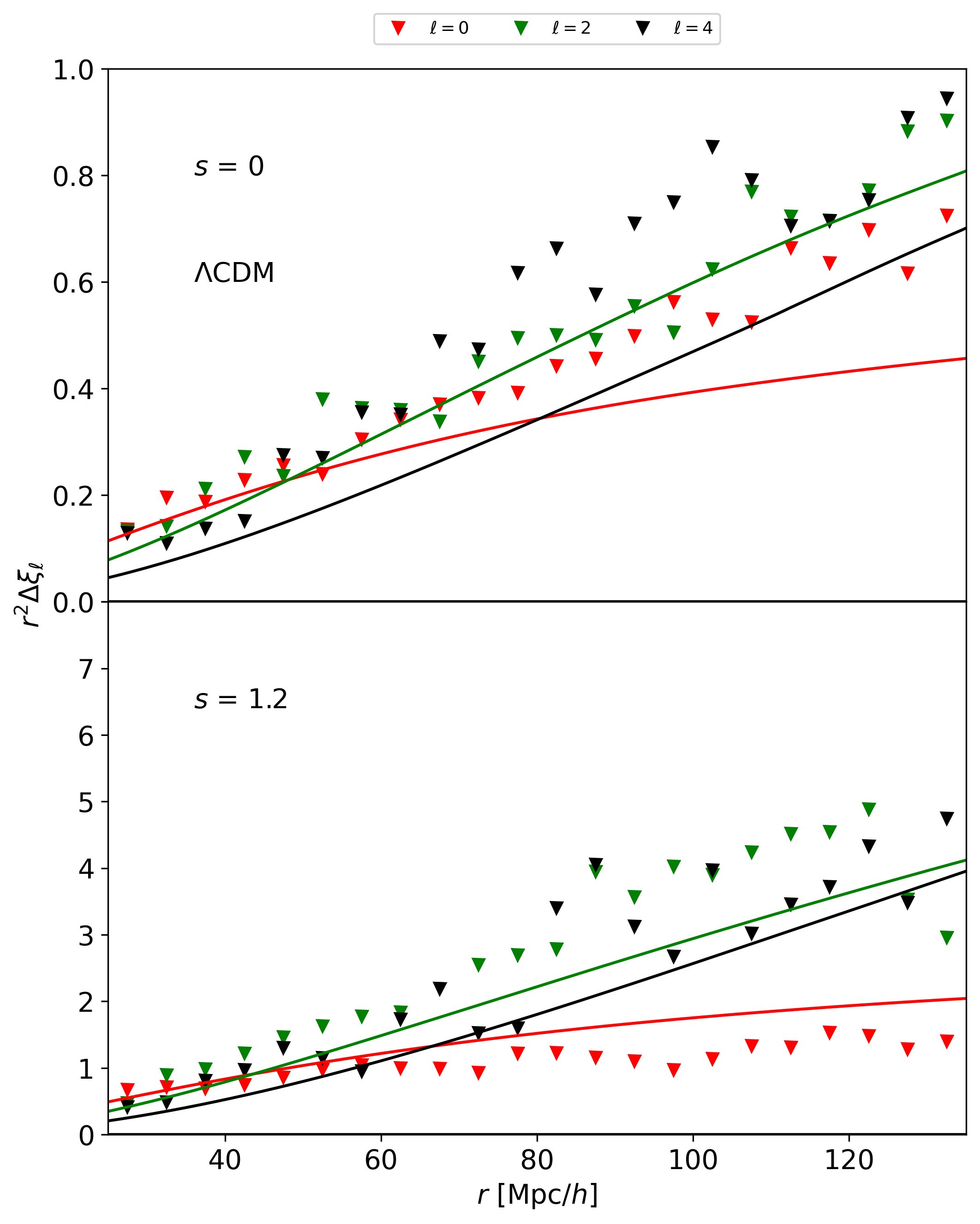}
    \includegraphics[width=\columnwidth]{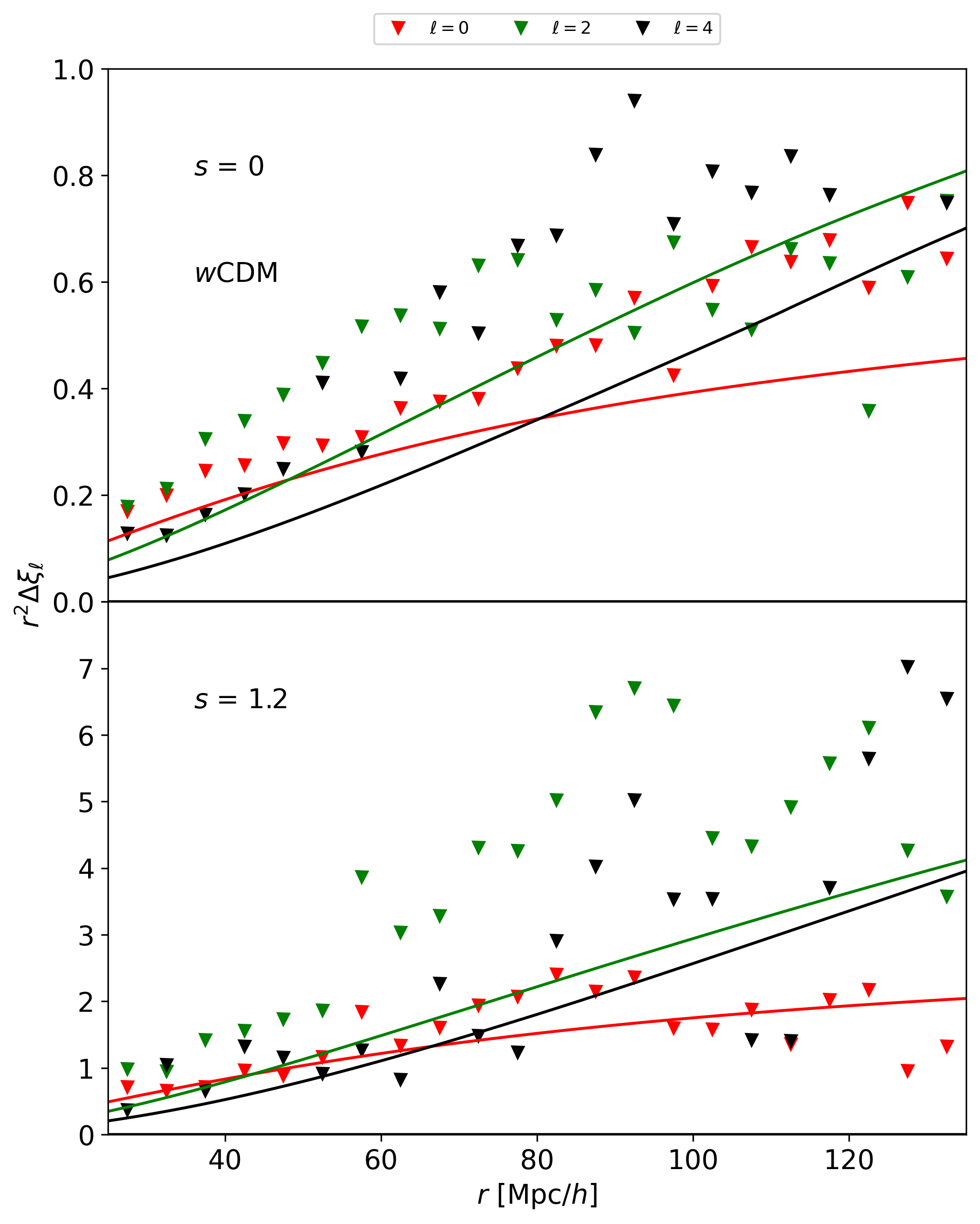}    
    \caption{Absolute difference in the correlation function multipoles for DM particles at the highest redshift bin (i.e. $z = 1.6-1.95$) when accounting for magnification bias with respect to the case with RSD only (left). The red and green points refer to the monopole and quadrupole in $\Lambda$CDM (left) and $w$CDM (right), while the red, green, and black lines show the $\Lambda$CDM theoretical prediction for the monopole, quadrupole, and hexadecapole computed with \textsc{Coffe}}
    \label{fig:absdiff_multipoles_lensing}
\end{figure*}

In this appendix we provide more details about the effect of the magnification bias on the multipole moments of the correlation function. We consider the DM particles in the highest studied redshift bin at about $z=1.8$, which maximises the magnification bias signal. The magnification bias corrections on the monopole, quadrupole, and hexadecapole of the correlation function are shown in \cref{fig:absdiff_multipoles_lensing}. 
First, we note that in any case, that is, for $s = 0$ or 1.2, $\Lambda$CDM or $w$CDM, the lensing contribution to the correlation function is always positive at this redshift. This is interesting because it confirms the results of \cref{fig:multipoles_theoretical} and explains why, if lensing magnification is not incorporated in the model, the likelihood analysis tries to compensate for this lack by increasing the value of the bias parameter (mostly due to the monopole) and therefore lower the value of the growth rate (as described in \cref{sec:results}) due to the quadrupole.

Secondly, for the case with $s = 0$, that is when we use observed angles instead of comoving ones, the monopole seems to agree with the theoretical prediction up to $60~h^{-1}$Mpc only. This is surprising as we would expect a good agreement at large scales. This difference might be due to the large variance inherent to these scales or to an inaccuracy in the modelling as well as the non-linearity of lensing corrections (see also \citealt{hui2007anisotropic} for a discussion). We also note the remarkable agreement between data and prediction for the quadrupole (while the hexadecapole seems underestimated in the prediction). 

Thirdly, focusing on the case with $s = 1.2,$ which give the most significant trend, the theoretical prediction overestimates the monopole and underestimates the quadrupole and hexadecapole above 80~$h^{-1}$Mpc. While this discrepancy is not necessarily large, especially for this high value of $s$, it impacts the estimation of the growth rate (see \cref{fig:fs8_lcdm_part}). This suggests that we should be careful about the first-order solution given within the weak-lensing limit, as higher-order terms might need to be accounted for at higher redshifts for high values of $s$.

Lastly, the theoretical prediction in the fiducial $\Lambda$CDM model does not agree very well with the data from the $w$CDM simulation in any case. Although the shape is similar between data points and analytical prediction, the amplitude is different (this difference is clear for the quadrupole, where data points are consistently at least a factor two higher than the prediction for both $s = 1.2$).
For more precise studies at higher redshift, it will be important to find a way to account for this cosmology-dependant correction. Nonetheless, even if the present modelling in the fiducial $\Lambda$CDM cosmology is not perfect for analysing the $w$CDM, it is still much better than not accounting for lensing magnification at all.

\end{document}